\documentclass[twocolumn,aps,superscriptaddress,showpacs,nofootinbib,floatfix]{revtex4-1}
\usepackage{epsfig,bm,feynmf}
\usepackage{graphicx}
\usepackage{amsmath}
\usepackage{dcolumn}

\usepackage{hyperref}
\hypersetup{
  colorlinks,
  unicode=True,
  linkcolor=Blue,
  citecolor=Blue,
  urlcolor=Blue,
}

\usepackage{graphicx}
\usepackage[usenames,dvipsnames,svgnames,table]{xcolor}

\usepackage{tikz-feynman,contour}
\tikzfeynmanset{compat=1.0.0}
\usepackage[normalem]{ulem}  

\renewcommand\sout{\bgroup \color{red} \ULdepth=-.5ex \ULset}

\begin{document}

\title{The influence of heavy quark potential on quarkonium production in quark-gluon plasma}

\author{Taesoo Song}\email{t.song@gsi.de}
\affiliation{GSI Helmholtzzentrum f\"{u}r Schwerionenforschung GmbH, Planckstrasse 1, 64291 Darmstadt, Germany}

\author{Jiaxing Zhao}\email{jzhao@itp.uni-frankfurt.de}
\affiliation{Institute for Theoretical Physics, Johann Wolfgang Goethe Universit\"{a}t, Frankfurt am Main, Germany}
\affiliation{Helmholtz Research Academy Hessen for FAIR (HFHF),GSI Helmholtz Center for Heavy Ion Research. Campus Frankfurt, 60438 Frankfurt, Germany}


    


\begin{abstract}
Remler formalism is the Wigner projection of a two particle state to bound states, which is carried out when the bound state begins to exist or when one of the two particles scatters in medium.
This method has been successfully applied to quarkonium production in box simulations and in heavy-ion collisions.
In this study this method is extended to strongly bound states with considerable binding energies by taking into account the potential energy between heavy quark pairs in the quark-gluon plasma (QGP).
We find that an attractive heavy-quark potential enhances quarkonium production and the results are consistent with those from the statistical model in box simulations, if a proper spatial cutoff is introduced to the potential to mimic quantum effects.
\end{abstract}


\maketitle

\section{Introduction}
Quarkonium production in heavy-ion collisions has intensively been studied since the seminal study of Matsui and Satz who proposed the $J/\psi$ suppression as a signature of the onset of quark-gluon plasma formation~\cite{Matsui:1986dk}.  
It, however, turned out that the story is more complicated, because the regeneration of $J/\psi$ from charm quark pairs becomes more and more important as collision energy increases and even prevails $J/\psi$ suppression from the Debye screening and/or thermal dissociation at LHC energies~\cite{ALICE:2019nrq,ALICE:2023hou}.
Therefore it is necessary for a model to include both quarkonium suppression and regeneration in heavy-ion collisions~\cite{Andronic:2006ky,Grandchamp:2002wp,Yan:2006ve,Song:2011xi,Yao:2020xzw,Brambilla:2016wgg,Blaizot:2018oev,Villar:2022sbv,Ferreiro:2018wbd}.
Models quite depend on the strength of binding of quarkonium.
If the binding of quarkonium is weak in QGP, most quarkonium will be formed near $T_c$ and the statistical hadronization model will be the best approach for it~\cite{Andronic:2006ky}.
In the case of relatively strong binding even in QGP, two-component model will be more reasonable to describe it correctly~\cite{Grandchamp:2002wp,Yan:2006ve}.

In our previous studies the Remler formalism~\cite{Remler:1975re,Remler:1975fm,Remler:1981du} was tested in a thermalized box~\cite{Song:2023ywt} and applied to quarkonium production in pp and heavy-ion collisions~\cite{Song:2017phm,Zhao:2023dvk,Song:2023zma}.
The Remler formalism is nothing but the projection of two particle state into one bound state which is carried out initially at the formation time and then whenever interaction of one of the two particles takes place.
Since the update of Wigner projection at scattering implies discarding old projection and taking new one, it effectively includes both gain and loss terms, which can respectively be interpreted as the regeneration and thermal decay of quarkonium from heavy quark scattering~\cite{Villar:2022sbv,Song:2023ywt}. 

It has been shown that in a thermalized box the Wigner projection produces the same number of bound state as in the statistical model~\cite{Song:2023ywt}. For example, 
\begin{eqnarray}
N&=&\int\frac{d^3p_1 d^3r_1}{(2\pi)^3} \int\frac{d^3p_2 d^3r_2}{(2\pi)^3} e^{-E_1/T} e^{-E_2/T}W(r,p) \nonumber\\
&\approx& V\int\frac{d^3P}{(2\pi)^3} e^{-E/T},
\label{loosely}
\end{eqnarray}
where $E_1=m+p_1^2/(2m)$, $E_2=m+p_2^2/(2m)$ with $m_1=m_2=m$. $W(r,p)$ is the Wigner function with $r=r_1-r_2$ and $p=(p_1-p_2)/2$, and $E=M+P^2/(2M)$ with $M=2m$ and $P=p_1+p_2$~\cite{Song:2023ywt}.
It explains why the coalescence model is successful for a loosely bound system such as deuteron whose mass is almost twice nucleon mass due to the small binding energy.
We note that the same assumption is made in our previous study for $J/\psi$ production in a box by using the Remler formalism~\cite{Song:2023ywt}.

However, if the binding energy is not small, that is, $M > M_\Phi$ with $M_\Phi$ being the mass of bound state, Eq.~(\ref{loosely}) will underestimate the number density of $\Phi$ in (grand) canonical ensemble by the factor $e^{-\varepsilon_0/T}$:
\begin{eqnarray}
(2\pi)^3\frac{dN}{d^3Rd^3P}\approx  e^{-(M_\Phi+P^2/2M)/T}e^{-\varepsilon_0/T}.
\label{benergy}
\end{eqnarray}
where $\varepsilon_0$ is the positive binding energy of $\Phi$ ($M=M_\Phi+\varepsilon_0$).

The statistical hadronization model well describes the experimental data on $J/\psi$ production in heavy-ion collisions, especially at LHC where charm quarks are closer to thermal equilibrium than at lower energies~\cite{Andronic:2019wva,ALICE:2023hou}.
The initial hard scatterings in heavy-ion collisions at high energies normally produce more charm quarks than in chemical equilibrium, and the annihilation of charm quark pair does not sufficiently happen during the short life time of QGP~\cite{Song:2024hvv,Song:2024afz}.
In order to consider the chemical off-equilibrium of charm quark, the charm fugacity, $\gamma_c$ is introduced in the statistical model such that
\begin{eqnarray}
N_{\rm open}^{Exp.}&=&\gamma_c N_{\rm open}^{Eq.}(T_{\rm cf}), \nonumber\\
N_{\rm hidden}^{Exp.}&=&\gamma_c^2 N_{\rm hidden}^{Eq.}(T_{\rm cf}), 
\label{gammac}
\end{eqnarray}
where $N_{\rm open}^{Exp.}$ and $N_{\rm hidden}^{Exp.}$ are the numbers of open and hidden charms measured in experiments and $N_{\rm open}^{Eq.}$ and $N_{\rm open}^{Eq.}$ are the numbers of open and hidden charms in the statistical model at the chemical freeze-out temperature ($T_{\rm cf}$) under the assumption of chemical equilibrium. For example,
\begin{eqnarray}
N_{\rm open}^{Eq.}(T_{\rm cf})=V\sum_i d_i\int \frac{d^3p}{(2\pi)^3} e^{-\sqrt{m_i^2+p^2}/T_{\rm cf}},
\end{eqnarray}
where $i$ sums over all existing physical states with the spin degeneracy factor $d_i$ and mass $m_i$, and $V$ is the volume of matter at the temperature of $T_{\rm cf}$.

On the other hand, charm density in QGP in the grand canonical ensemble is given by 
\begin{eqnarray}
N_{\rm open}^{\rm QGP}(T,m_c)=V d_c\int \frac{d^3p}{(2\pi)^3} e^{-\sqrt{m_c^2+p^2}/T},
\label{charm-number}
\end{eqnarray}
where the color-spin degeneracy factor of charm $d_c=6$ and $m_c$ is the effective charm quark mass.
In our recent study $m_c \approx$ 1.8 GeV at $T_c \approx$ 0.16 GeV is required for charm density to be smoothly connected from QGP to hadron gas phase, which really takes place in heavy-ion collisions~\cite{Song:2024rjh}.
We note that the similar results are found in Refs.~\cite{Grandchamp:2003uw,Zhao:2010nk,Riek:2010fk,Liu:2017qah}.

As shown in Eq.~(\ref{loosely}), in order for the coalescence model or the Remler formalism, both of which use the Wigner projection, to correctly estimate $J/\psi$ production from charm quark and anticharm quark in equilibrium, the relation, $m_{J/\psi}\approx 2 m_c$, should be satisfied, which means $m_c\approx$ 1.55 GeV.
However, it requires much more (anti)charm quarks about by factor 4:
\begin{eqnarray}
\frac{N_{\rm open}^{\rm QGP}(T_c,m_c=1.55~ {\rm GeV})}{N_{\rm open}^{\rm QGP}(T_c,m_c=1.8~ {\rm GeV})}=4.
\end{eqnarray}
Therefore, the coalescence model (or the Wigner projection) will underestimate $J/\psi$ production, if it is carried out near $T_c$.

In principle, Wigner function which is used in the coalesecne model and Remler formalism is constructed from a wavefunction which needs an attractive potential.
However, our previous studies assumed free heavy (anti)quarks without potential in QGP for their dynamics~\cite{Song:2023ywt,Song:2023zma}.
Although the heavy quark potential have been taken into account for heavy quark dynamics in Ref.~\cite{Villar:2022sbv}, the color combination of heavy quark pairs is not systematically treated.
For example, if a heavy quark or heavy antiquark interacts with a thermal parton, 
the color combination of the heavy quark pair changes from color singlet to color octet or from color octet to color singlet or to another color octet.

In this study we investigate the effect of heavy quark potential on quarkonium production in the Remler formalism.
Since the attractive potential draws heavy quark near heavy antiquark, it will compensate the underestimate of quarkonium production caused by the binding energy in Eq.~(\ref{benergy}). 
As a first step we test this approach in thermalized box and compare the results with those in the statistical model.

This paper is organized as follows:
Remler formalism is briefly reviewed in Sec.~\ref{remler} and applied to quarkonium, after discussing about two-body system in sections~\ref{pairs}  and~\ref{quarkonium} where we also find a problem which arises from semi-classical approach.
This problem is investigated in two examples in sections~\ref{1dsho} and \ref{test2}, and then a solution to it is presented with the results on quarkonium in Sec.~\ref{quantum}.
Finally summary is given in Sec.~\ref{summary}.

\section{Remler formalism}\label{remler}

Now let us briefly discuss about its impact on the Remler formalism for quarkonium production in QGP~\cite{Remler:1975re,Remler:1975fm,Remler:1981du}.
In the Remler formalism the number of quarkonium is counted by the Wigner projection of quarkonium into heavy quark and heavy antiquark distribution~\cite{Song:2017phm,Zhao:2023dvk} and the projection is updated whenever heavy quark or heavy antiquark interacts and changes it position and momentum~\cite{Villar:2022sbv,Song:2023zma}.  
For example, the expected number of quarkonium is given by 
\begin{equation}
    N(t\to\infty) = N(0)+\int_0^\infty \Gamma(t)dt,
\label{probability}    
\end{equation}
where the rate $\Gamma(t)$ is defined as 
\begin{eqnarray}
\Gamma(t)
&=& \sum_{i,j}\sum_{\nu_{i(j)}}\frac{1}{(2\pi)^{3N}}  \int d^3r_1d^3p_1 ... d^3r_N d^3p_N\nonumber\\
&\times& W_0(\vec{r}_i-\vec{r}_j,\vec{p}_i-\vec{p}_j)
\bigg\{\delta\bigg(t-t^\nu_{i(j)}\bigg)\nonumber\\
&-&\delta\bigg(t-t^{\nu-1}_{i(j)}\bigg)\bigg\}W^{(N)}(t+\varepsilon),
\label{new}
\end{eqnarray}
where the first summation of $i$ and $j$ on the right hand side respectively runs over all heavy quarks and antiquarks, $t^\nu_{i(j)}$ is the time of the $\nu$' th scattering of the heavy quark $i$ (or of the heavy antiquark $j$) with a QGP parton, $W_0$ is the Wigner function of quarkonium 
which is nothing but the Wigner transformation of the density operator of pure quarkonium state:
\begin{eqnarray}
W_0(\vec{r},\vec{p})&=&\int d^3r^\prime\bigg\langle \vec{r}-\frac{\vec{r^\prime}}{2}\bigg|\hat{\rho}\bigg|\vec{r}+\frac{\vec{r^\prime}}{2}\bigg\rangle \exp\bigg\{i\vec{p}\cdot\vec{r}^\prime\bigg\}\nonumber\\
&=&\int d^3r^\prime e^{i\vec{p}\cdot \vec{r}^\prime}\psi\bigg(\vec{r}+\frac{\vec{r}^\prime}{2}\bigg)\psi^*\bigg(\vec{r}-\frac{\vec{r}^\prime}{2}\bigg),
\label{wigner-general}
\end{eqnarray}
where $\hat{\rho}=|\Phi\rangle\langle\Phi|$ with $\Phi$ being quarkonium and $\psi(r)=\langle \Phi |\vec{r}\rangle$, while $W^{(N)}(t)=W^{(N)}(\vec{r}_1,\vec{p}_1,\vec{r}_2,\vec{p}_2,...,\vec{r}_N,\vec{p}_N;t)$ is quantal density matrix in Wigner representation of the $N$ partons which is normalized as  
\begin{eqnarray}
\int d^3r_1d^3p_1 ... d^3r_N d^3p_N W^{(N)}(t)=(2\pi)^{3N}
\label{normalization0}
\end{eqnarray}
regardless of $t$. Since the quantal density matrix is not in general known, it is replaced by an average of classical phase space density distribution given as
\begin{eqnarray}
W^{(N)}(t)\approx \prod_{i=1}^N (2\pi)^{3N}\delta(r_i-r_i^*(t))\delta(p_i-p_i^*(t)).
\label{deltas}
\end{eqnarray} 
$\varepsilon$ in Eq.~(\ref{new}) implies a slight time delay to take into account the momentum change of the heavy (anti)quark due to the scattering with a QGP parton at $t=t^\nu_{i(j)}$ or $t=t^{\nu-1}_{i(j)}$. 
The first and second terms in the curly bracket of Eq.~(\ref{new}), respectively, indicate the gain and loss terms of quarkonium.

The initial Wigner projection in Eq.~(\ref{probability}) is, based on Eq.~(\ref{new}), given by
\begin{eqnarray}
N(0)
&=& \sum_{i,j}\frac{1}{(2\pi)^{3N}}  \int d^3r_1d^3p_1 ... d^3r_N d^3p_N\nonumber\\
&\times& W_0(\vec{r}_i-\vec{r}_j,\vec{p}_i-\vec{p}_j)
W^{(N)}(0),
\label{begin}
\end{eqnarray}
where $t=0$ means the initial formation time of quarkonium. 
This form can be understood from Eq.~(\ref{new}), considering that 
quarkonium begins to exist below its dissociation temperature $(T_d)$ in heavy-ion collisions.
In other words, the temperature at $t=t^\nu_{i(j)}$ is below 
$T_d$ and $W_0$ exists, while the temperature at $t=t^{\nu-1}_{i(j)}$ is above $T_d$ and $W_0$ does not exists. 
Therefore, the loss term vanishes and 
only the gain term remains, which is equivalent to Eq.~(\ref{begin}).

Integrating the rate over all time, Eq.~(\ref{probability}) turns to 
\begin{eqnarray}
N(t\to\infty)&=&\sum_{i,j}\frac{1}{(2\pi)^{3N}}  \int d^3r_1d^3p_1 ... d^3r_N d^3p_N\nonumber\\
&\times& W_0(\vec{r}_i-\vec{r}_j,\vec{p}_i-\vec{p}_j)\bigg\{
W^{(N)}(0)\nonumber\\
&+&W^{(N)}(t^1_{i(j)})-W^{(N)}(0)+~...\nonumber\\
&+&W^{(N)}(t^f_{i(j)})-W^{(N)}(t^{f-1}_{i(j)})\bigg\}\nonumber\\
&=&\sum_{i,j}\frac{1}{(2\pi)^{3N}}  \int d^3r_1d^3p_1 ... d^3r_N d^3p_N\nonumber\\
&\times& W_0(\vec{r}_i-\vec{r}_j,\vec{p}_i-\vec{p}_j)W^{(N)}(t^f_{i(j)}),
\label{tinfinity}
\end{eqnarray}
where $t^f_{i(j)}$ is the last scattering time of heavy quark $i$ or heavy antiquark $j$.
This is nothing but the Wigner projection of $ W_0(\vec{r},\vec{p})$ into $W^{(N)}(t)$ for each heavy quark pair combination at $t=t^f_{i(j)}$.

In thermal equilibrium, the ensemble average of $N-$body density matrix is same at any time:
\begin{eqnarray}
\langle W^{(N)}(0)\rangle= \langle W^{(N)}(t^1_{i(j)})\rangle = ... =\langle W^{(N)}(t^f_{i(j)})\rangle.
\end{eqnarray}
In other words, $N(t)$ in Eq.~(\ref{probability}) does not change with time, because the matter is thermalized.
Ignoring the potential between heavy quark and heavy antiquark and treating them as free particles, their distribution in phase space will be like the Boltzmann distribution in non-relativistic limit,
\begin{eqnarray}
\sum_{i,j}\frac{1}{(2\pi)^{3(N-2)}}  \int d^3r_3d^3p_3 ... d^3r_N d^3p_N W^{(N)}\nonumber\\\Rightarrow d_1e^{-(m_1+p_1^2/2m_1)/T}d_2e^{-(m_2+p_2^2/2m_2)/T},
\end{eqnarray}
where $d_1$ and $d_2$ are respectively color and spin degeneracy of heavy quark and heavy antiquark.
Then Eq.~(\ref{tinfinity}) is simplified into  
\begin{eqnarray}
N(t\to\infty)&=&\frac{d_1d_2}{(2\pi)^{6}}  \int d^3r_1d^3p_1 d^3r_2 d^3p_2\nonumber\\
&\times& W_0(\vec{r}_1-\vec{r}_2,\vec{p}_1-\vec{p}_2)e^{-(m_1+p_1^2/2m_1)/T}\nonumber\\
&\times& e^{-(m_2+p_2^2/2m_2)/T},
\label{coal}
\end{eqnarray}
where $W_0$ should be multiplied by the degeneracy factors $D/(d_1d_2)$ with $D$ being the spin degeneracy of quarkonium to match the physical quantum numbers of quarkonium.
We note that Eq.~(\ref{coal}) is quite similar to so-called coalescence model for quarkonium formation in QGP.

This approach has been successful in describing bottomum production not only in pp and but also heavy-ion collisions at LHC energies~\cite{Song:2023zma}.
However, the interaction or potential between heavy quark and heavy antiquark may affect their dynamics in heavy-ion collisions, though it will be less important for bottom quarks compared to charm quarks~\cite{Villar:2022sbv}.
In this study we extend our previous study by including the heavy quark potential in the Remler formalism, but as a first step it is considered in a thermalized box which is easy to control. 

\section{heavy quark pair}\label{pairs}

The partition function for two-body system  in canonical ensemble is given by
\begin{eqnarray}
Z(T,V,2)=\int \frac{dr_1^3 dp_1^3}{(2\pi)^3}
\int \frac{dr_2^3 dp_2^3}{(2\pi)^3}e^{-\beta H(r_1,p_1;r_2,p_2)},
\end{eqnarray}
with $\beta\equiv 1/T$. Since we have in mind a heavy quark pair, the Gibb's factor $2!$ is not divided.
Now we think about the partition function for $N$ heavy quark pairs:
\begin{eqnarray}
&&Z(T,V,2N)=\nonumber\\
&&\frac{1}{N!}\bigg(\int \frac{dr_1^3 dp_1^3}{(2\pi)^3}\int \frac{dr_2^3 dp_2^3}{(2\pi)^3}e^{-\beta H(r_1,p_1;r_2,p_2)}\bigg)^N,
\end{eqnarray}
where $1/N!$ is the Gibb's factor.
Here we assume that a heavy quark affects only one heavy antiquark and vice versa.  
It will not be a bad approximation if heavy quark density is low such that the interspace between heavy quark and antiquark is larger than the range of heavy quark potential.

Then the grand canonical partition function is given by
\begin{eqnarray}
&&{\it Z}(T,V,\mu)=\sum_n g_c^nZ(T,V,2n)\\
&&=\exp\bigg(\int \frac{dr_1^3 dp_1^3}{(2\pi)^3}
\int \frac{dr_2^3 dp_2^3}{(2\pi)^3}e^{-\beta H(r_1,p_1;r_2,p_2)-\beta \mu_c}\bigg)\nonumber
\end{eqnarray}
where the heavy quark fugacity $g_c=e^{-\beta \mu_c}$ with $\mu_c$ being heavy quark chemical potential, and 
the number density of heavy quark pair by
 
\begin{eqnarray}
&&N(T,V,\mu)=-\frac{\partial \Phi}{\partial \mu_c}\nonumber\\
&&=\int \frac{dr_1^3 dp_1^3}{(2\pi)^3}
\int \frac{dr_2^3 dp_2^3}{(2\pi)^3}e^{-\beta H(r_1,p_1;r_2,p_2)-\beta\mu_c},
\label{ccbarden}
\end{eqnarray}
where the grand canonical potential $\Phi=-T\ln{\it Z}(T,V,
\mu_c)$.

Hamiltonian of two body system in Eq.~(\ref{ccbarden}) can be rearranged into
\begin{eqnarray}
H\approx M+\frac{P^2}{2M}+\frac{p^2}{2\mu}+V(r),
\end{eqnarray}
where
\begin{eqnarray}
\vec{P}&=&\vec{p}_1+\vec{p}_2,~~~
\vec{p}=\frac{m_2\vec{p}_1-m_1\vec{p}_2}{m_1+m_2},\nonumber\\
\vec{R}&=&\frac{m_1\vec{r}_1+m_2\vec{m}_2}{m_1+m_2},~~~
\vec{r}=\vec{r}_1-\vec{r}_2
\label{change-variables}
\end{eqnarray}
with $M=m_1+m_2$ and $\mu=m_1 m_2/(m_1+m_2)$.
Then the number density of heavy quark pair from Eq.~(\ref{ccbarden}) is expressed as 
\begin{eqnarray}
(2\pi)^3\frac{dN}{d^3Rd^3P}&=&g_c d_1d_2e^{-(M+P^2/2M)/T}\nonumber\\
&\times&\int d^3r\int \frac{dp^3}{(2\pi)^3}e^{-p^2/(2\mu T)-V(r)/T}\nonumber\\
&=&g_c d_1d_2\bigg(\frac{\mu T}{2\pi}\bigg)^{3/2}e^{-(M+P^2/2M)/T}\nonumber\\
&\times &\int d^3r e^{-V(r)/T},
\label{dist1}
\end{eqnarray}
where $d_1$ and $d_2$ are the spin-color degeneracies of heavy quark and heavy antiquark, respectively. 

Since particle with the same quantum number is indistinguishable, a proper spatial integration range will be the inverse particle density such that 
\begin{eqnarray}
\frac{4}{3}\pi \mathcal{R}^3 \approx \bigg(g_c d_2 \int \frac{dp^3}{(2\pi)^3}e^{-\sqrt{m_2^2+p^2}/T}\bigg)^{-1},
\label{r-range}
\end{eqnarray}
where $\mathcal{R}$ is the upper limit of $r$ integration.
Heavy quark potential for color-singlet and color-octet in pQCD are respectively given by~\cite{Halzen:1984mc,Peskin:1979va}
\begin{eqnarray}
V_s(r)&=& -\frac{N_c^2-1}{2N_c}\frac{\alpha_s}{r},\nonumber\\
V_o(r)&=&\frac{1}{2N_c}\frac{\alpha_s}{r}.
\label{V-pQCD}
\end{eqnarray}

The potentials in Eq.~(\ref{V-pQCD}) diverges at $r=$0.
However, $\alpha_s$ should vanish as $r\rightarrow 0$.
This effect can simply be realized by modifying the potential to be saturated around inverse heavy quark mass, $r=1/m_Q$. 
This simple cut-off can also be understood from the uncertainty principle~\cite{Song:2017phm,Song:2023zma}.

Considering only the color degeneracy in $d_1$ and $d_2$ in Eq.~(\ref{dist1}), the probability for a pair to be color singlet is 1/9 and that to be color octet is 8/9. 
However, the potential of color-singlet is attractive, while that of color-octet is repulsive, which enhances the probability to be color singlet and suppresses the probability for color octet through the potential $V(r)$ in Eq.~(\ref{dist1}).

\begin{figure} [h!]
    \includegraphics[width=8.5 cm]{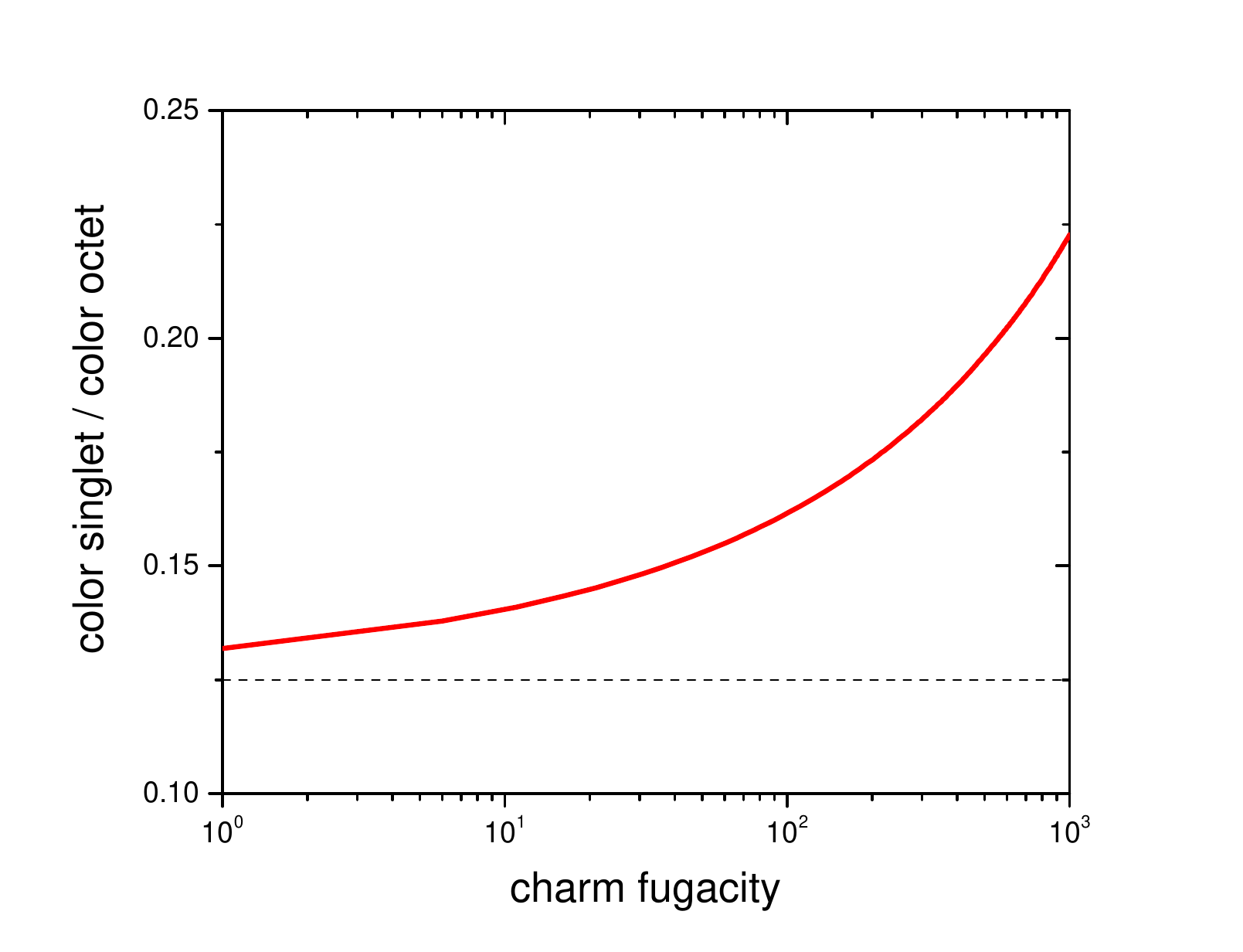} 
    \caption{The ratio of color singlet to color octet at $T_c$ as a function of charm fugacity in the presence of pQCD heavy quark potential from Eq.~(\ref{V-pQCD}), assuming $\alpha_s$ and $m_c$ are respectively $\pi/12$ and 1.8 GeV~\cite{Song:2024rjh}.}
    \label{color-singlet}
\end{figure}

Fig.~\ref{color-singlet} is the ratio of the number density of charm quark pairs in color-singlet to that in color-octet at $T_c=$ 0.158 GeV as a function of charm fugacity. That is, from Eqs.~(\ref{dist1}) and (\ref{r-range})
\begin{eqnarray}
\int^\mathcal{R}_{1/m_c}drr^2 e^{-V_s(r)/T}\bigg/
\int^\mathcal{R}_{1/m_c}drr^2e^{-V_o(r)/T}.
\end{eqnarray}
$\alpha_s$ and $m_c$ are respectively taken to be $\pi/12$ and 1.8 GeV~\cite{Song:2024rjh}.

Charm fugacity being one in Fig.~\ref{color-singlet} means charm quark number is chemically thermalized, and a larger charm fugacity implies more population of charm quark pairs than in chemical equilibrium.
One can see that even in equilibrium the number of color-singlets is larger than 1/8 which is indicated by the dashed line in the figure, and the ratio increases with increasing charm quark density, because the upper limit of spatial integration in Eq.~(\ref{dist1}), $\mathcal{R}$, decreases with increasing charm fugacity.
We, however, note that this is a very naive estimate, because in reality there are interactions between two charm quarks as well as between two anticharm quarks.

\section{quarkonium}\label{quarkonium}
In order to study the number of a bound state in the Remler formalism, the corresponding Wigner function is needed.
For example, simply assuming the wavefunction from the potential of simple harmonic oscillator, the Wigner function for 1S quarkonium state is given by 
\begin{eqnarray}
W_{1S}=8\frac{D}{d_1d_2}e^{-r^2/\sigma^2-p^2\sigma^2}
\label{wigner}
\end{eqnarray}
where $D$ is the spin degeneracy of 1S bound state, $d_1$ and $d_2$ the spin-color degeneracy factors of quark and antiquark and $\sigma^2=(2/3)\langle (r_1-r_2)^2 \rangle$.
Inserting Eq.~(\ref{wigner}) into Eq.~(\ref{dist1}) in order to obtain the number density of 1S state, 
\begin{eqnarray}
(2\pi)^3\frac{dN^{1S}}{d^3Rd^3P}&=&De^{-(M+P^2/2M)/T}\nonumber\\
&\times& \int \frac{dp^3}{\pi^3}e^{-p^2/(2\mu T)-p^2\sigma^2}\nonumber\\
&\times& \int d^3r e^{-V(r)/T-r^2/\sigma^2}.
\label{dist2}
\end{eqnarray}

If particles are extremely heavy ($\mu \rightarrow \infty$) and potential is negligible ($V(r)\rightarrow 0$), Eq.~(\ref{dist2}) simply turns to a Boltzmann distribution of 1S bound state:
\begin{eqnarray}
(2\pi)^3\frac{dN^{1S}}{d^3Rd^3P}\approx De^{-(M+P^2/2M)/T}.
\label{dist2-approx}
\end{eqnarray}

This is why the coalescence model is successful for a loosely bound system such as deutron whose mass is almost same as the sum of two nucleon mass ($M = m_1+m_2$) due to the small binding energy.

However, if the binding energy is not small, that is, $M > M_\Phi$ with $M_\Phi$ being the mass of bound state, Eq.~(\ref{dist2-approx}) will underestimate the number density of $\Phi$ by the factor of $e^{-\varepsilon_0/T}$ in (grand) canonical ensemble:
\begin{eqnarray}
(2\pi)^3\frac{dN^{1S}}{d^3Rd^3P}\approx D e^{-(M_\Phi+P^2/2M)/T}e^{-\varepsilon_0/T}.
\label{be}
\end{eqnarray}
where $\varepsilon_0$ is the positive binding energy of quarkonium ($M=M_\Phi+\varepsilon_0)$.
We note that $M$ in the denominator of $P^2/2M$ also should be changed to $M_\Phi$.
Additionally, $e^{-p^2/(2\mu T)}$ in Eq.~(\ref{dist2}) will further suppress the number density of $\Phi$, if $\mu$ is not sufficiently large.

These two suppression may be compensated by the attractive potential through $e^{-V(r)/T}$ in Eq.~(\ref{dist2}). 
In fact, the Wigner function in Eq.~(\ref{wigner}) is made of wavefunction of $\Phi$, for which the attractive potential is necessary.
Now we study the effects of the potential on quarkonium coalescence by considering the heavy quark potential energy from lattice QCD~\cite{Gubler:2020hft,Satz:2005hx}.

The calculations start with solving the Schr\"{o}dinger
equation
\begin{eqnarray}
\bigg[2m_0-\frac{\nabla^2}{m_0}+V(r,T)\bigg]\psi(r,T)=M_\Phi(T) \psi(r,T),
\end{eqnarray}
where $m_0$ is the bare mass of heavy quark which is fitted to reproduces quarkonium mass in vacuum (T=0).
It is taken to be 1.25 GeV for charm and 4.65 GeV for bottom~\cite{Andronic:2024oxz}.

Since the potential energy does not converge to 0 in the limit of $r\rightarrow \infty$, it is redefined by subtracting the potential energies at infinity, and heavy quark effective mass is redefined as
\begin{eqnarray}
\tilde{V}(r,T)&=&V(r,T)-V(r=\infty,T),\nonumber\\
m_Q(T) &=& m_0 + \frac{1}{2}V(r=\infty,T),
\label{mass0}
\end{eqnarray}
such that the Schr\"{o}dinger equation is modified into
\begin{eqnarray}
\bigg[-\frac{\nabla^2}{m_Q(T)}+\tilde{V}(r,T)\bigg]\psi(r,T)=-\varepsilon_0(T) \psi(r,T),
\label{Schrodinger}
\end{eqnarray}
where the (positive) binding energy $\varepsilon_0=2m_Q-M_\Phi$.
We note that the denominator of the first term is changed from $m_0$ to $m_Q$ by hand for the simplicity of calculations.

\begin{figure} [h!]
    \includegraphics[width=9.2 cm]{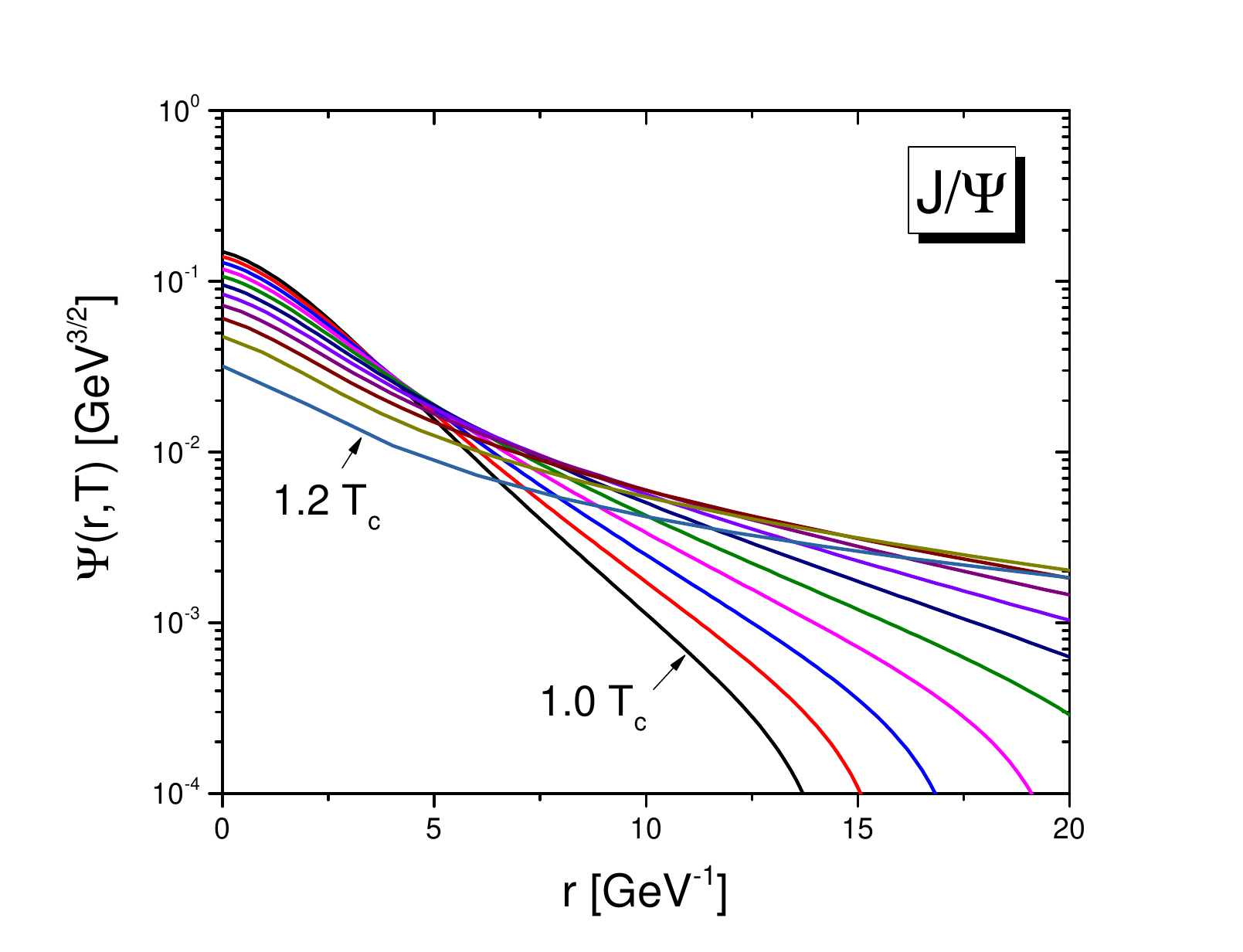} 
    \includegraphics[width=9.2 cm]{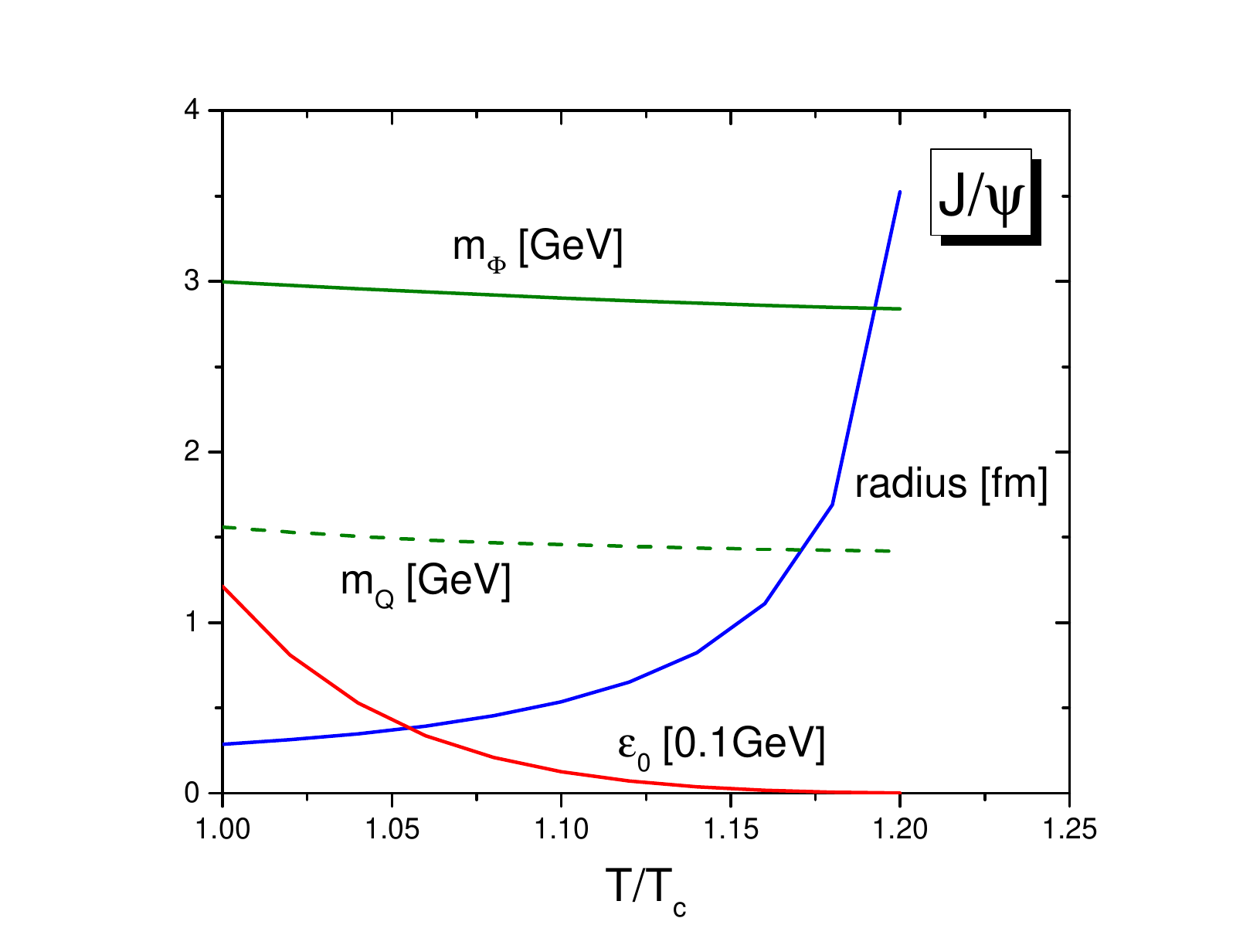}   
    \caption{(Upper) Wavefunction of $J/\psi$ at the temperature from 1.0 $T_c$ to 1.2 $T_c$ and (lower) the mass, binding energy and radius of $J/\psi$ along with charm quark mass for the free-energy potential~\cite{Gubler:2020hft,Satz:2005hx} as a function of temperature.}
    \label{schreodinger}
\end{figure}

Fig.~\ref{schreodinger} shows the numerical solutions of Eq.~(\ref{Schrodinger}) with the free-energy potential~\cite{Gubler:2020hft,Satz:2005hx}, that is, the wavefunction and binding energy which correspond respectively to the eigenfunction and eigenvalue, along with the average radius, charm quark mass and $J/\psi$ mass as a function of temperature.

\begin{figure} [h!]
    \includegraphics[width=8.5 cm]{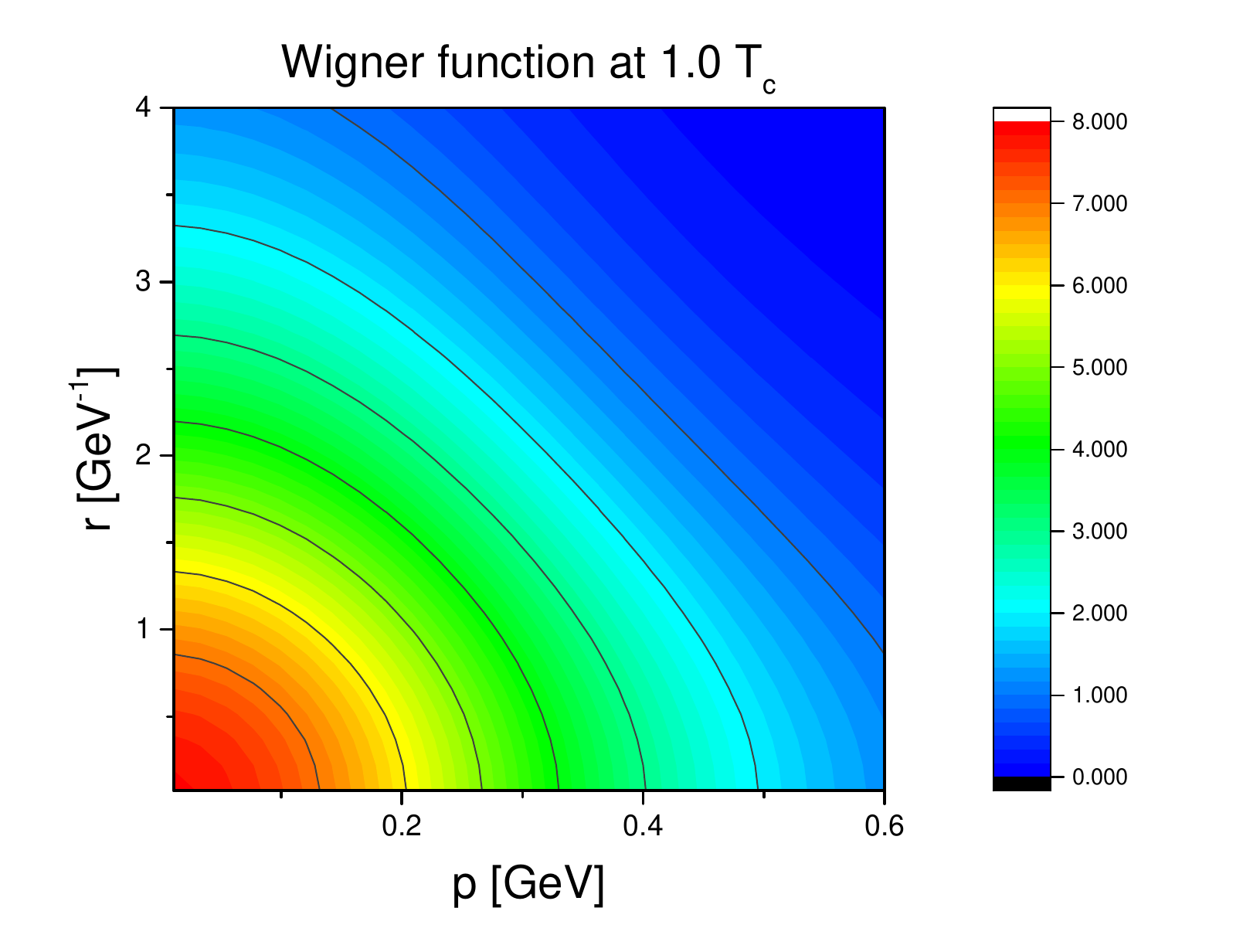} 
    \includegraphics[width=8.5 cm]{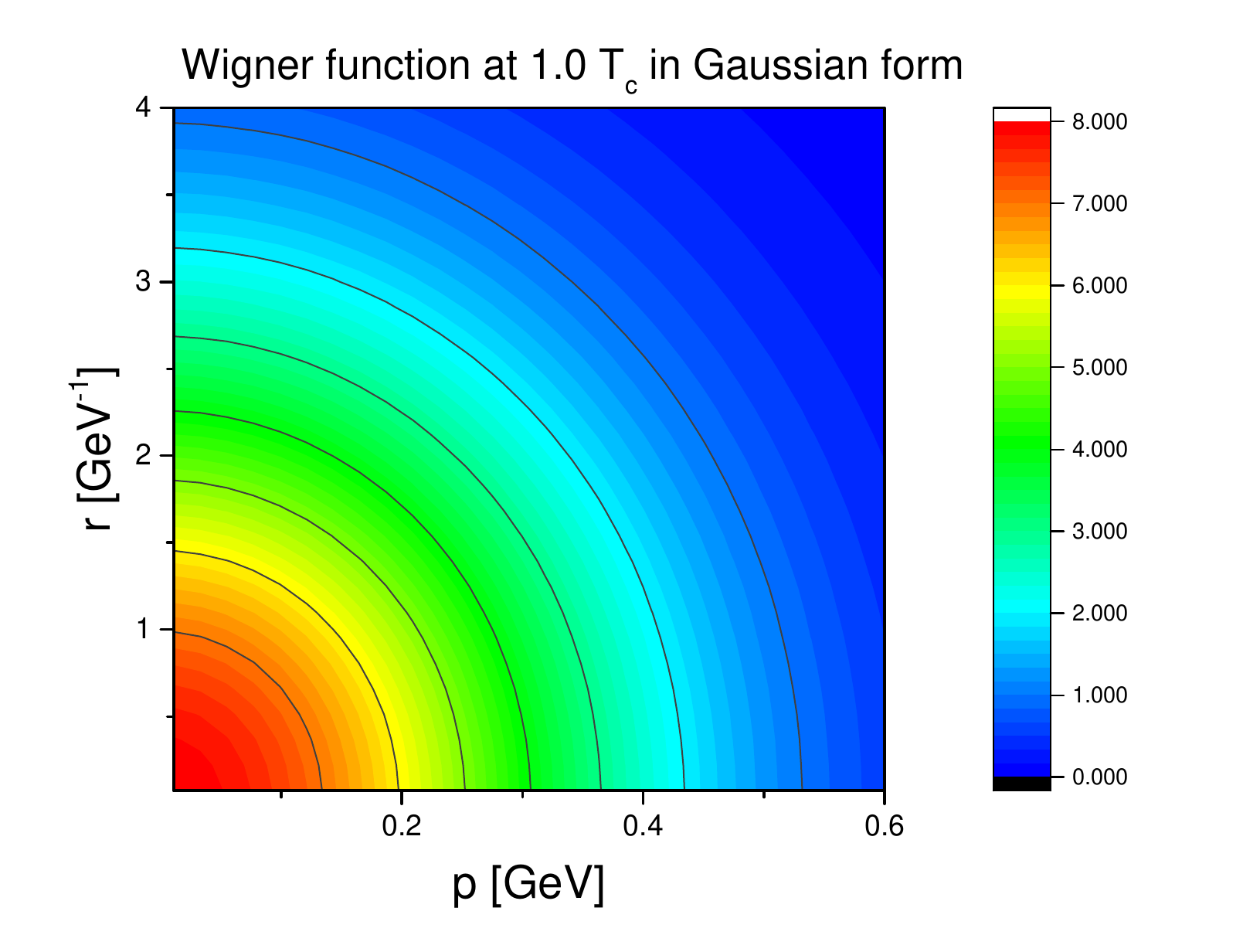}     
    \caption{Wigner functions at 1.0 $T_c$ (upper) from the definition of the Wigner function in Eq.~(\ref{wigner-general}) and (lower) from the gaussian form in Eq.~(\ref{wigner}).}
    \label{sho}
\end{figure}

Then the Wigner function, defined in Eq.~(\ref{wigner-general}) but often simplified into Eq.~(\ref{wigner}), is obtained from the wavefunction.
Fig.~\ref{sho} shows the Wigner function from its definition and from the gaussian form, which correspond respectively to Eqs.~(\ref{wigner-general}) and (\ref{wigner}).
We note that in the figures the angle between $\vec{r}$ and $\vec{p}$ is integrated.
One can see that both Wigner functions are similar to each other and satisfy the normalization condition:
\begin{eqnarray}
\int d^3r d^3p W_0(\vec{r},\vec{p})=(2\pi)^3.
\label{normalization}
\end{eqnarray}

Now we compare $J/\psi$ productions from the Wigner density projection and from the statistical model. For simplicity, their ratio at $P=0$ is given by 
\begin{eqnarray}
&&\bigg(\frac{dN^{Wig.}}
{d^3Rd^3P}\bigg|_{P=0}\bigg)\bigg/\bigg(\frac{dN^{stat.}}{d^3Rd^3P}\bigg|_{P=0}\bigg)\\
&&=e^{-\varepsilon_0/T}\int \frac{dp^3}{\pi^3}e^{-p^2/(2\mu T)-p^2\sigma^2}\int_{1/\mu} d^3r e^{-V(r)/T-{r^2\over \sigma^2}},\nonumber
\label{ratio}
\end{eqnarray}
from Eqs.~(\ref{dist2}) and (\ref{be}).
We note that the lower limit of $r$ integration is set at $1/\mu=2/m_Q$ to prevent the infrared divergence at $r=0$, which appears in the Coulomb-like potential ($V(r)\sim -1/r$):
\begin{eqnarray}
\int_{0} dr r^2e^{-V(r)/T}\rightarrow \infty.
\end{eqnarray}
Eq.~(\ref{ratio}) demonstrates again that if the potential is weak ($V(r)\rightarrow 0,~ \varepsilon_0 \rightarrow 0$) and the mass is large ($\mu \rightarrow \infty$), the ratio converges to unity. 

\begin{figure} [h!]
    \includegraphics[width=8.5 cm]{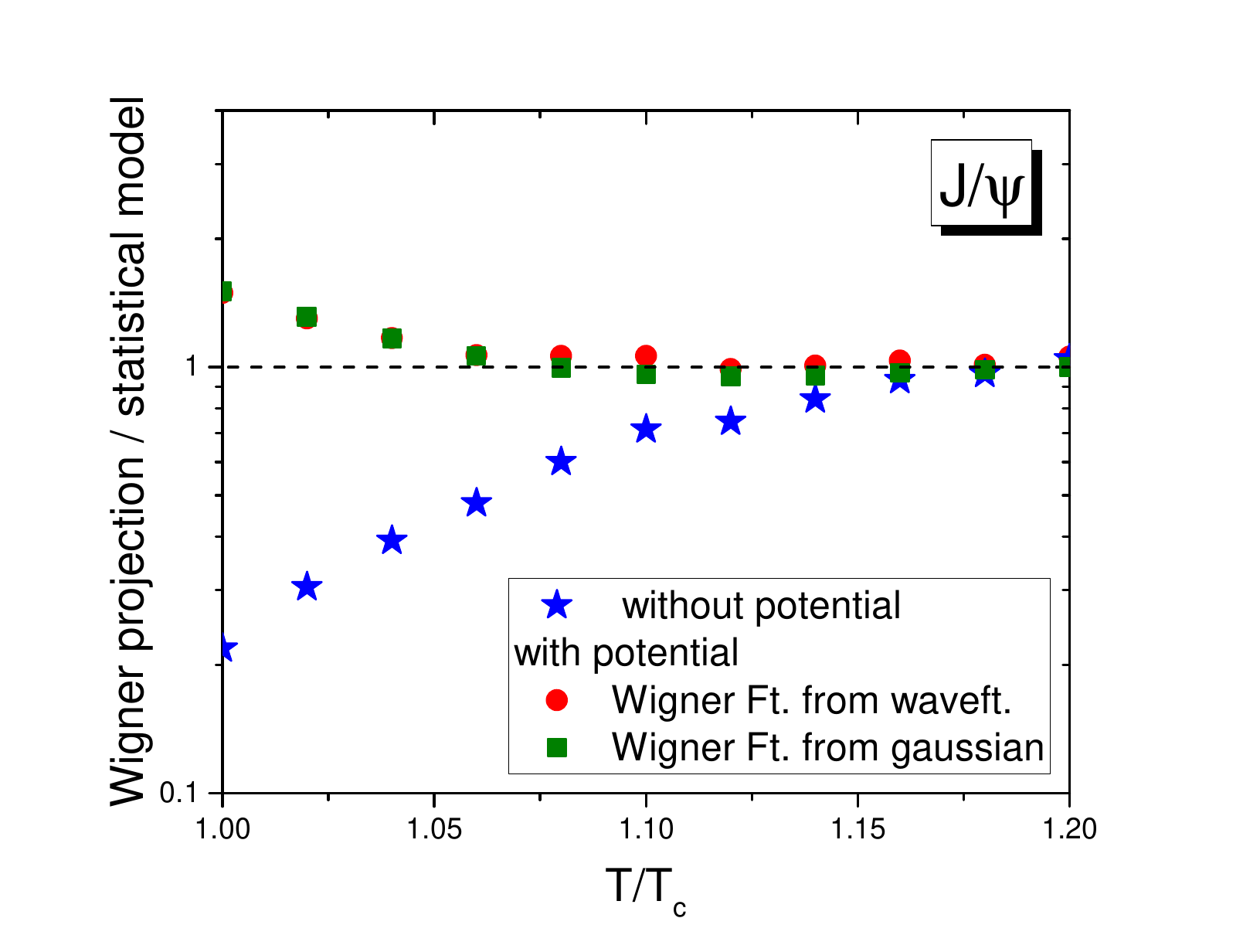} 
    \includegraphics[width=8.5 cm]{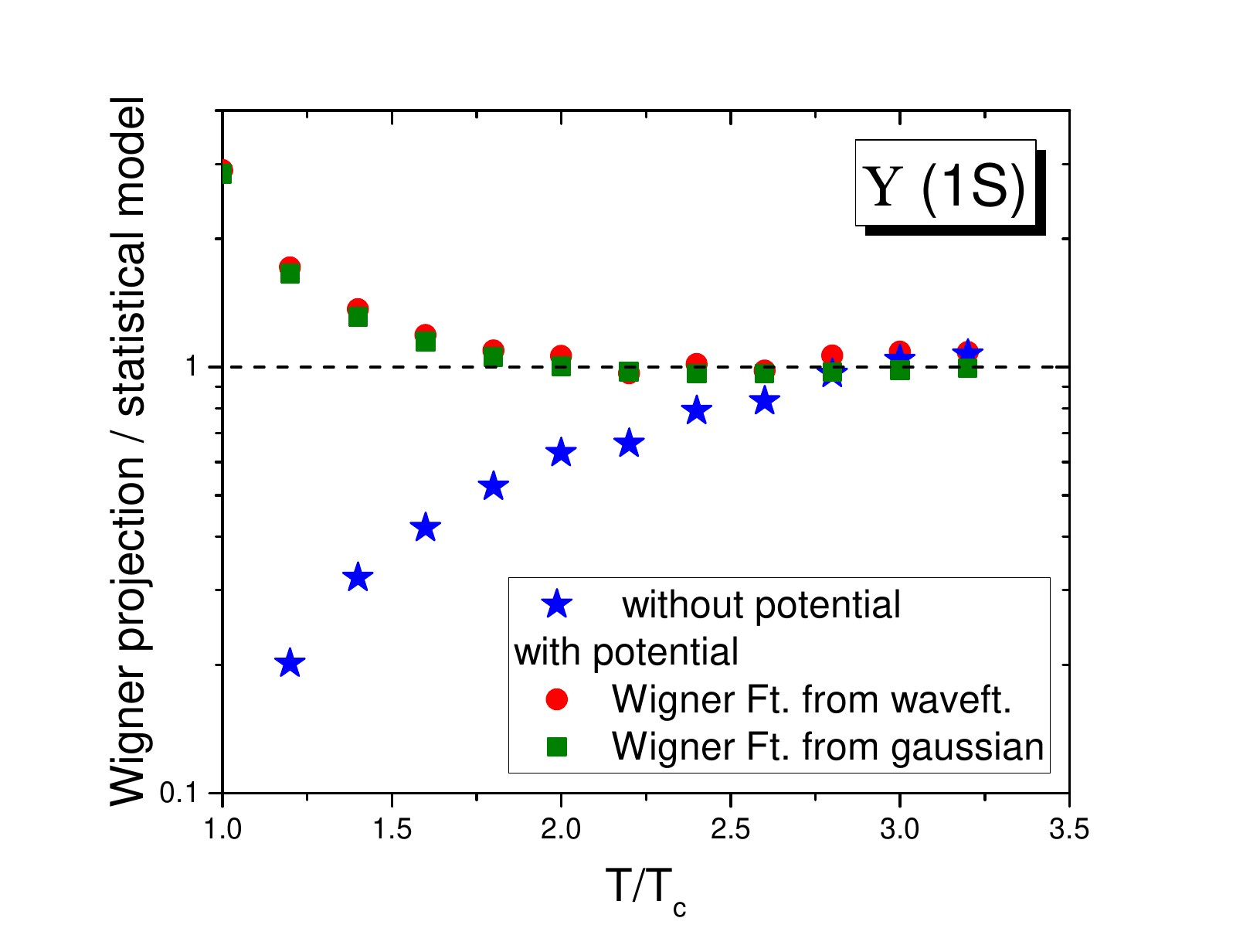}     
    \caption{The ratios of the number density from the Wigner projection to that from the statistical model for (upper) $J/\psi$ and (lower) $\Upsilon (1S)$ with and without the modification of the heavy quark distribution due to the free energy heavy-quark potential~\cite{Gubler:2020hft,Satz:2005hx}. The ratio is separated into the two different cases where the Wigner function is calculated directly from the wavefunction as in Eq.~(\ref{wigner-general}) and calculated from the gaussian form in Eq.~(\ref{wigner}).}
    \label{compare}
\end{figure}

Fig.~\ref{compare} displays the ratios of the number density from the Wigner projection to that from the statistical model for $J/\psi$ and $\Upsilon (1S)$ as a function of rescaled temperature.
The blue stars indicate the ratio without considering the potential in the heavy quark distribution in coordinate space, in other words, without $\exp[-V(r)/T]$ in Eq.~(\ref{ratio}).
One can see that the number densities of both $J/\psi$ and $\Upsilon (1S)$ in the Remler formalism are similar to those in the statistical model when the binding energy is small or temperature is close to the dissociation temperature.
However, the ratios begin to decrease with decreasing temperature, partly because of the increasing binding energy and partly due to the term $\exp[-p^2/(2\mu T)]$ in Eq.~(\ref{ratio}).
The ratio drops down to 0.22 for $J/\psi$ and to 3 \% for $\Upsilon (1S)$ at $T_c$.

However, the low ratios are recovered by including the heavy quark potential which attracts heavy quark and heavy antiquark to each other. 
The red circles and green squares in Fig.~\ref{compare} are the ratios after including heavy quark potential.
The color combination of heavy quark pair can be either color singlet or color octet.
But we ignore the color octet state, because it cannot form a bound state.
The inclusion of the heavy quark potential really helps the Wigner projection to reproduce the statistical results in thermal equilibrium, but it begins to overestimate as temperature approaches $T_c$.
The ratio of Wigner projection to statistical model reaches 1.5 for $J/\psi$ and even 3.0 for $\Upsilon (1S)$ at $T_c$.

In the next two sections, we present the reason for the overestimation of Wigner projection at low temperature 
through the examples of 1-dimensional simple harmonic oscillator and of the screened Coulomb potential.

\section{1-dimensional simple harmonic oscillator}\label{1dsho}

The potential for simple harmonic oscillator is $V(x)=(K/2)x^2$ with $K$ being the string tension and the particle distribution at temperature $T$ is given by 
\begin{eqnarray}
f(x,p)=A\exp\bigg[-\frac{p^2/(2m)+(K/2)x^2}{T}\bigg],
\end{eqnarray}
where $A$ is the normalization factor such that 
\begin{eqnarray}
1=\int \frac{dxdp}{2\pi}f(x,p)=AT\sqrt{\frac{m}{K}}.
\end{eqnarray}
That is,
\begin{eqnarray}
f(x,p)=\sqrt{\frac{K}{mT^2}}\exp\bigg[-\frac{p^2+mKx^2}{2mT}\bigg].
\end{eqnarray}

On the other hand, the Wigner function for 1-dimensional harmonic oscillator is given by
\begin{eqnarray}
W_n(x,p)=\int dx^\prime e^{ipx^\prime}\psi_n\bigg(x+\frac{x^\prime}{2}\bigg)\psi_n^*\bigg(x-\frac{x^\prime}{2}\bigg),
\label{wigner-1dim}
\end{eqnarray}
where $\psi_n(x)$ is the wavefunction with $n$'th excited state:
\begin{eqnarray}
\psi_n(x)=\bigg(\frac{1}{\pi\sigma^2}\bigg)^{1/4}\frac{H_n(x/\sigma)}{\sqrt{2^n n!}}e^{-x^2/(2\sigma^2)}
\end{eqnarray}
with $H_n$ being the Hermite polynomial and $\sigma^2=1/\sqrt{mK}$.
For example, the ground state
\begin{eqnarray}
W_0(x,p)=2e^{-x^2/\sigma^2-\sigma^2 p^2}.
\label{wigner0}
\end{eqnarray}

Applying the Remler formalism, the probability for the ground state is given by 
\begin{eqnarray}
P_0=\int \frac{dxdp}{2\pi}f(x,p)W_0(x,p)=\frac{2}{1+2 T/\omega},
\label{prob-wg}
\end{eqnarray}
where $\omega=\sqrt{K/m}$.
However, Eq.~(\ref{prob-wg}) is different from the probability in quantum statistics~\cite{Greiner:1995vqa}:
\begin{eqnarray}
P_0=\frac{1}{Z}e^{-\omega/(2T)}=1-e^{-\omega/T},
\label{qs}
\end{eqnarray}
where $Z=[2\sinh(\omega/(2T))]^{-1}$ is the partition function.

\begin{figure} [h!]
    \includegraphics[width=8.5 cm]{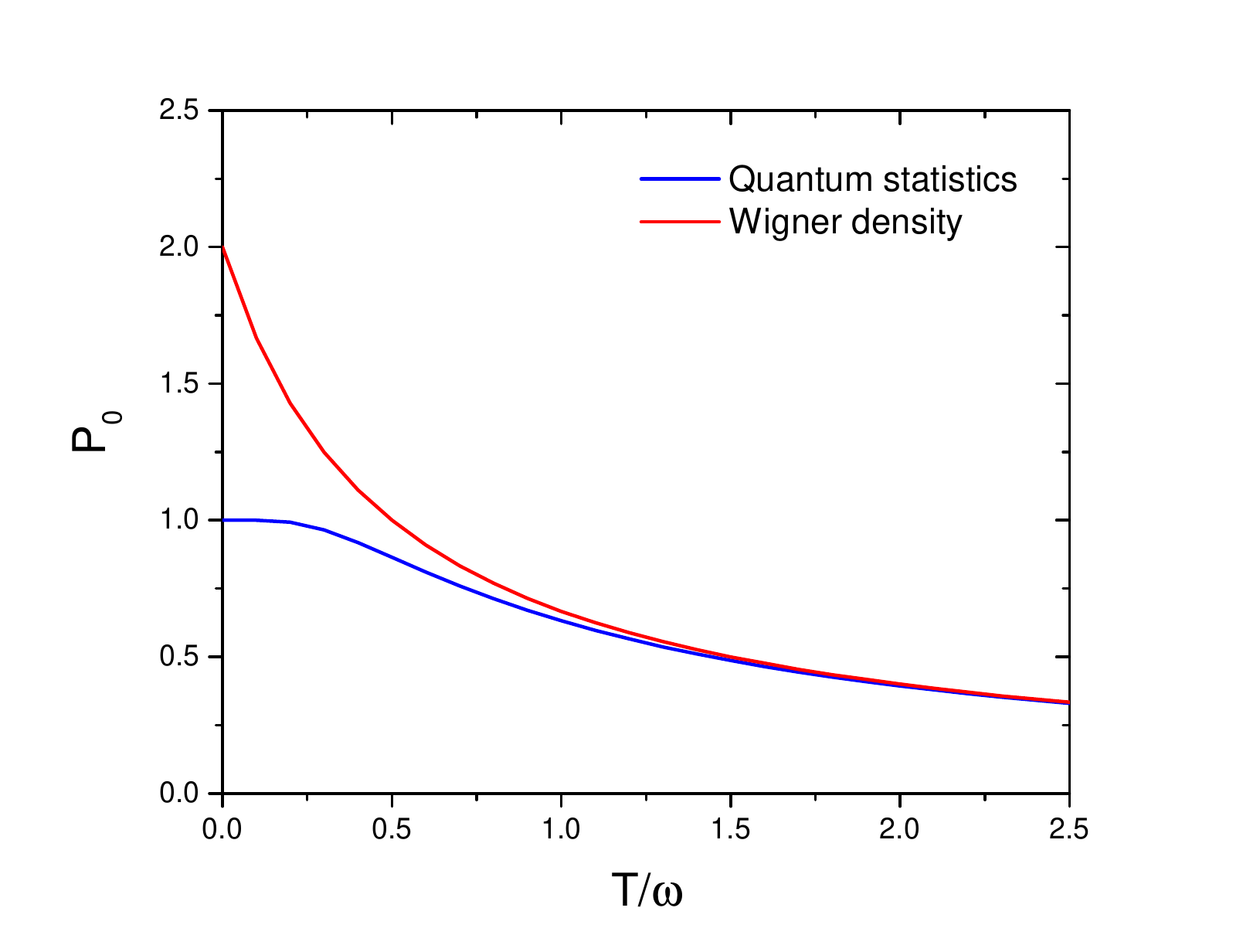}     
    \caption{The probability for the ground state in 1-dimensional simple harmonic oscillator as a function of scaled temperature in quantum statistics and from the Wigner projection}
    \label{qm-wigner}
\end{figure}

Fig.~\ref{qm-wigner} shows the probability for the ground state in 1-dimensional simple harmonic oscillator in quantum mechanics and from the Wigner projection.
In the low temperature limit ($T\rightarrow 0$) the probability for the ground state from the Wigner projection is 2 ($P_0= 2$).
It is understandable from Eq.~(\ref{wigner0}), because 
$x,p\rightarrow 0$ as $T\rightarrow 0$, but a probability larger than 1 is nonphysical.
On the other hand, $P_0$ from the Wigner projection is consistent with that from quantum statistics as temperature increases, and both Eq.~(\ref{prob-wg}) and Eq.~(\ref{qs}) converge to $\omega/T$ at high temperature.

In quantum mechanics position and momentum cannot be specified simultaneously due to the uncertainty principle.
For example, the off-diagonal component of the density operator of free gas in volume $V$ is given by~\cite{Greiner:1995vqa}
\begin{eqnarray}
\langle \vec{r^\prime} | \rho | \vec{r}\rangle =\frac{1}{V}\exp\bigg[-\frac{\pi}{\lambda^2}(\vec{r^\prime}-\vec{r})^2\bigg],
\end{eqnarray}
where the thermal wavelength $\lambda=\sqrt{2\pi/mT}$ which is infinitely large at $T=0$ but negligible at high temperature.
Furthermore, the density distribution of an oscillator in the ground state at $T=0$ is not a delta function $\delta(\vec{r})$ but is given by~\cite{Greiner:1995vqa}
\begin{eqnarray}
\langle \vec{r} | \rho | \vec{r}\rangle =\frac{1}{\sqrt{\pi}\sigma}\exp\bigg[-\frac{r^2}{\sigma^2}\bigg].
\end{eqnarray}

\section{the screened Coulomnbic potential}\label{test2}

The binding energy in Coulomnbic potential is given by
\begin{eqnarray}
E_n=-\frac{R}{n^2},
\end{eqnarray}
where the Rydberg constant $R=\hbar^2/(2\mu a_0^2)$ with $\mu$ being the reduced mass and $a_0$ being the Bohr radius (=$\hbar^2/\mu e^2$).
Considering the degeneracy factor for $n$' th excited
state is $n^2$, the partition function from the bound states 
\begin{eqnarray}
Z_{bound}=\sum_{n=1}^\infty n^2\exp\bigg(\frac{R}{n^2T}\bigg)
\label{bound}
\end{eqnarray}
and from unbound states
\begin{eqnarray}
Z_{unbound}&=&\sum_k \exp(-E_k/T)\nonumber\\
&=&\frac{V}{(2\pi)^3}\int d^3p\exp\bigg(-\frac{p^2}{2\mu T}\bigg)\nonumber\\
&=&V\bigg(\frac{\mu T}{2\pi}\bigg)^{3/2},
\label{unbound}
\end{eqnarray}
where $V$ is a volume.
Since the wavefunction size is infinitely large for $n\rightarrow \infty$ in Eq.~(\ref{bound}), $V$ in Eq.~(\ref{unbound}) diverges~\cite{Blinder:1995}.
Therefore, the probability for the ground state vanishes unless $T\rightarrow 0$:
\begin{eqnarray}
P_0=\frac{e^{R/T}}{Z_{bound}+Z_{unbound}}\rightarrow 0,~~~ {\rm for}~ T\ne 0.
\label{r}
\end{eqnarray}

However, as temperature increases, more and more electrons are set free, forming a QED plasma in which the Coulomb potential is screened~\cite{Bellac:2011kqa}:
\begin{eqnarray}
-\frac{4\pi\alpha}{r}\rightarrow -\frac{4\pi\alpha}{r}e^{-m_D r},
\end{eqnarray}
with the Debye screening mass $m_D=eT/\sqrt{3}$.
Then the number of bound states becomes finite and $V$ in Eq.~(\ref{unbound}) is limited due to the screening effect.
In other words, both $Z_{bound}$ and $Z_{unbound}$ become finite and $P_0$ will not vanish.
Since there is no analytic solution of wavefunction and eigenenergy for the screened potential, one should rely on numerical calculations.


Now we turn to the Remler formalism. Assuming that only one pair of particles exists in volume $V$, from Eq.~(\ref{dist1})
\begin{eqnarray}
P_0=\frac{4\pi}{A}\int_{1/\mu}^{r_D} dr r^2\int \frac{d^3p}{(2\pi)^3}e^{-(p^2/2\mu +V(r))/T}W_0(r,p),\nonumber\\
\label{p0real}
\end{eqnarray}
where $A$ is the normalization factor 
\begin{eqnarray}
A=4\pi \int_{1/\mu}^{r_D} dr r^2\int \frac{d^3p}{(2\pi)^3}e^{-(p^2/2\mu+V(r))/T}
\label{normalization}
\end{eqnarray}
with $r_D=(3V/4\pi)^{1/3}$.
We note that the lower limit of $r$ integration ($1/\mu$) is necessary to prevent the divergence at $r=0$.

In the limit of high temperature ($V(r)/T\rightarrow 0$), one can find the normalization factor of Eq.~(\ref{normalization}) is equivalent to the partition function for the unbound states in Eq.~(\ref{unbound}).
Furthermore, Eq.~(\ref{p0real}) is approximated as
\begin{eqnarray}
P_0&\approx&\frac{1}{Z_{unbound}}\int d^3r\int \frac{d^3p}{(2\pi)^3}e^{-p^2/(2\mu T)}W_0(r,p)\nonumber\\
&\approx& \frac{1}{Z_{unbound}}\int d^3r\int \frac{d^3p}{(2\pi)^3}W_0(r,p)\nonumber\\
&\approx& \frac{1}{Z_{unbound}}.
\label{approx}
\end{eqnarray}
The last approximation is valid if the expectation value of the bound state radius squared is large enough $\langle r^2\rangle \gg 1/\mu T$~\cite{Song:2023ywt}.
Considering that the binding energy of the ground state, $R$ in Eq.~(\ref{r}), must be very small at high temperature, one can claim that 
Eq.~(\ref{approx}) is equivalent to Eq.~(\ref{r}).

\begin{figure} [h!]
    \includegraphics[width=8.5 cm]{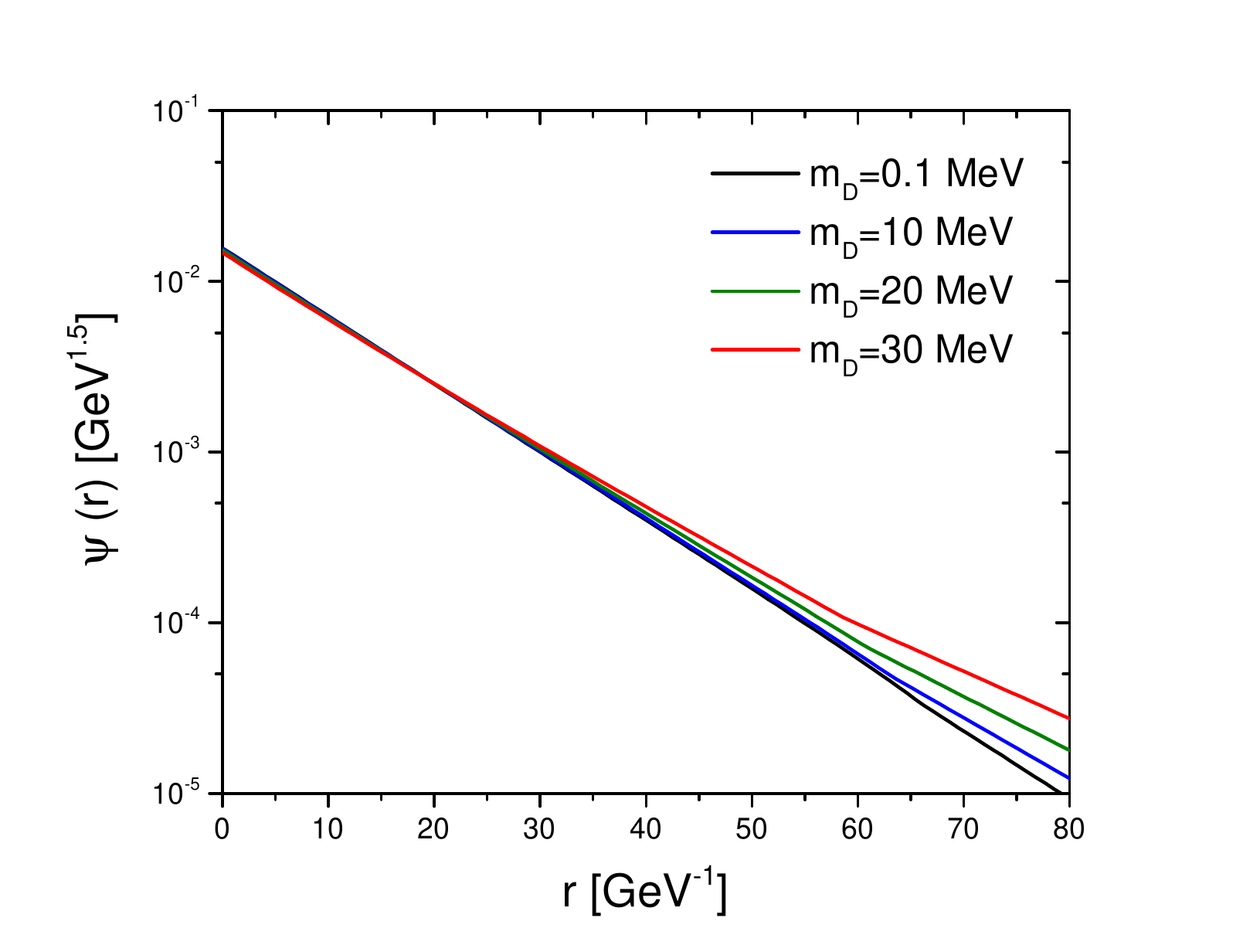}     
    \includegraphics[width=8.5 cm]{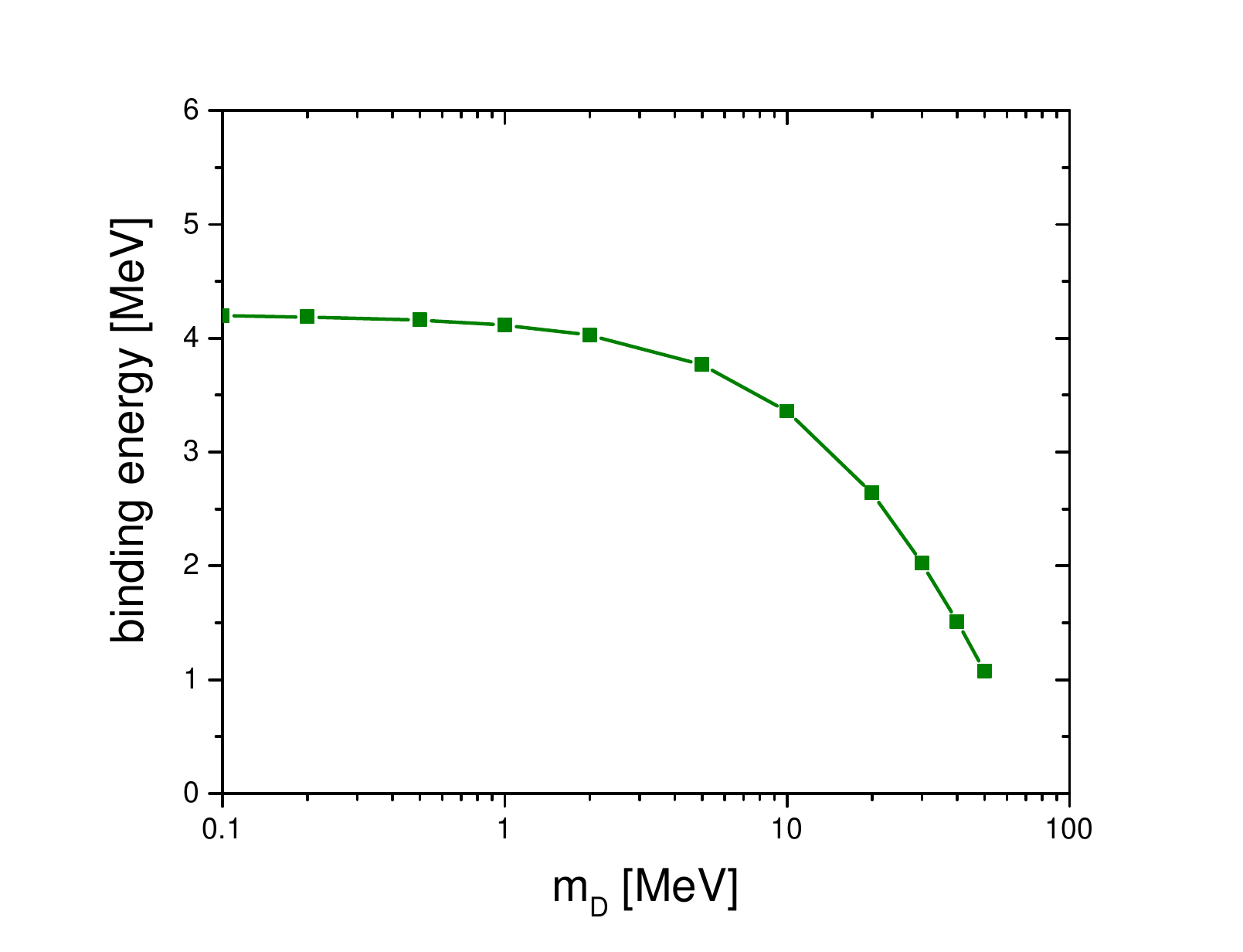}        
    \caption{(Upper) The wavefunctions and (lower) binding energies of the ground state (1S) from the screened Coulomb potential for several different screening masses. For simplicity, the reduced mass $\mu$ is taken to be 1 GeV.}
    \label{waveft}
\end{figure}

Fig.~\ref{waveft} shows the wavefunctions and binding energies of the ground state obtained by solving Schr\"{o}dinger equation with the screened Coulomb potential for several screening masses.
The reduced mass $\mu$ is taken to be 1 GeV for simplicity.
One can see that the wavefunction and binding energy are consistent with analytic solutions in vacuum ($R=$ 4.2 MeV at $m_D=0$ or $T=0)$ and rapidly change after $m_D=$ 10 MeV.

\begin{figure} [h!]
    \includegraphics[width=8.5 cm]{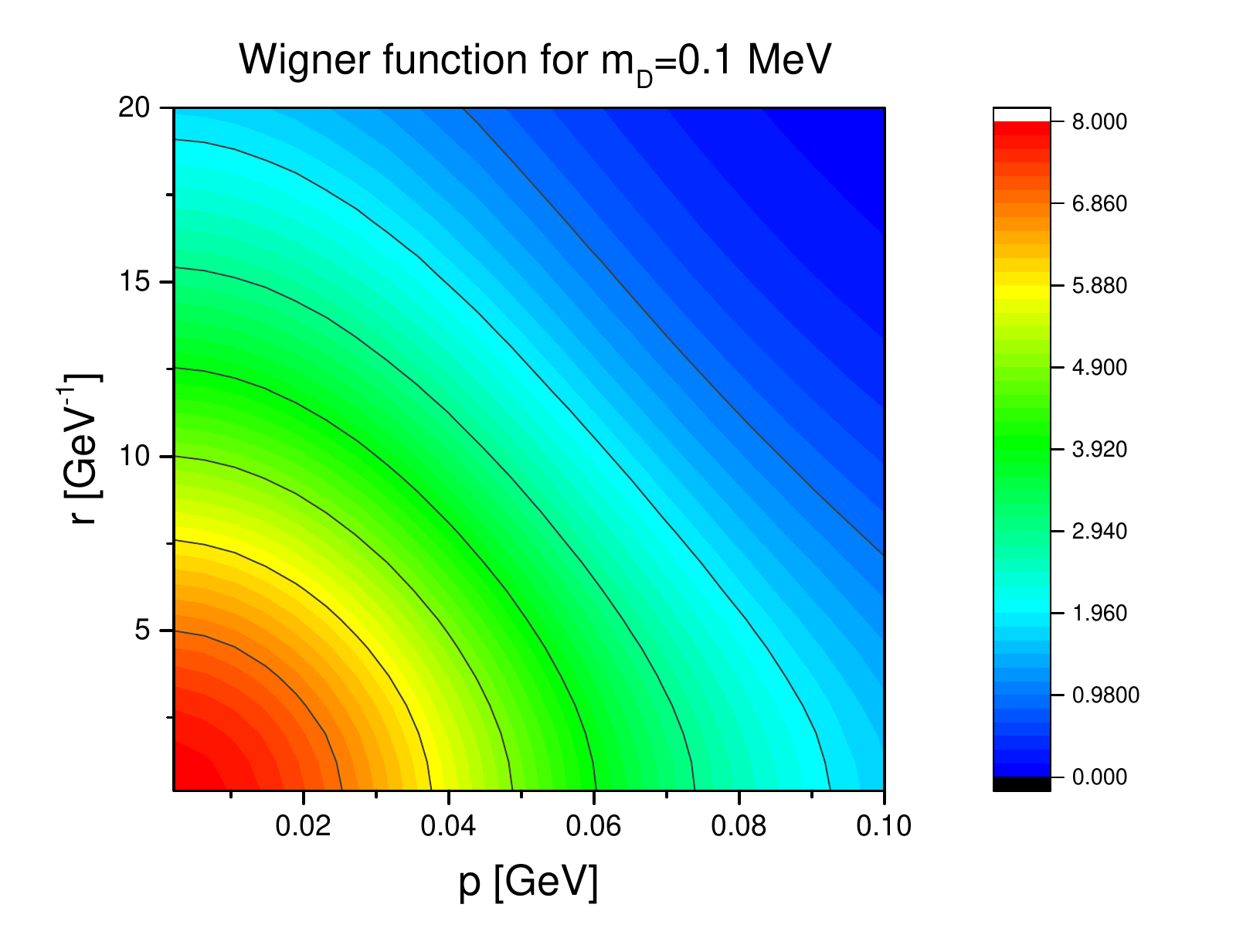}     
    \includegraphics[width=8.5 cm]{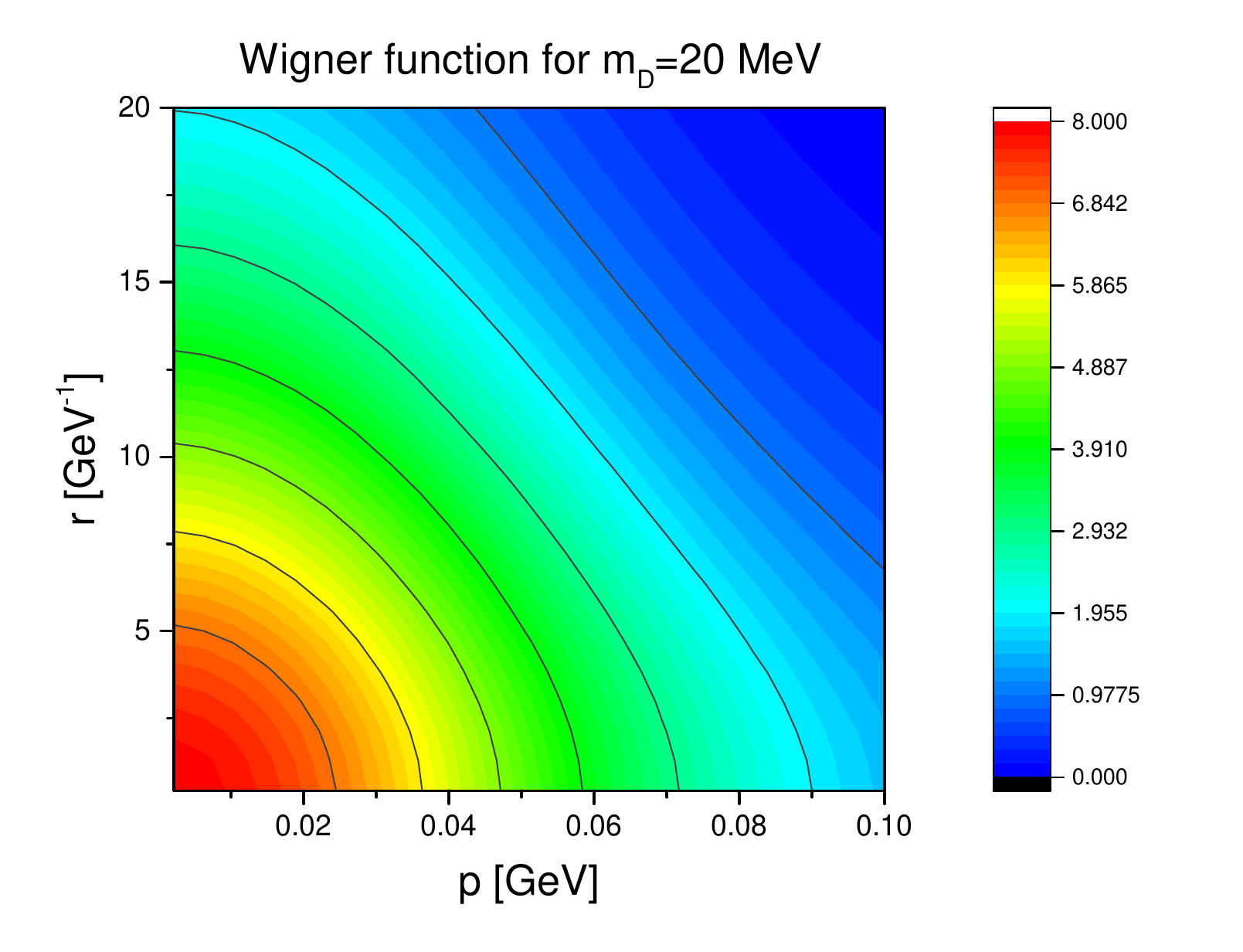}        
    \caption{Wigner function as a function of radial distance and momentum from the screened Coulomb potential for $m_D$=0.1 MeV and 20 MeV where the angle between $\vec{p}$ and $\vec{r}$ is averaged.}
    \label{wigner-fig}
\end{figure}

Then the Wigner function is calculated from the wavefunction which is similar to Eq.~(\ref{wigner-1dim}):
\begin{eqnarray}
W_0(\vec{r},\vec{p})=\int d^3r^\prime e^{i\vec{p}\cdot \vec{r}^\prime}\psi_0\bigg(\vec{r}+\frac{\vec{r}^\prime}{2}\bigg)\psi_0^*\bigg(\vec{r}-\frac{\vec{r}^\prime}{2}\bigg),
\label{wigner-3dim}
\end{eqnarray}
where $\psi_0 (\vec{r})$ is the ground state wavefunction.
We note that the Wigner function in general depends on the angle between $\vec{p}$ and $\vec{r}$ even for the ground state~\cite{Cho:2014xha}. 

Fig.~\ref{wigner-fig} shows the Wigner function as a function of radial distance and momentum for the screening mass $m_D$ of 0.1 MeV and of 20 MeV.
The angle between $\vec{p}$ and $\vec{r}$ is averaged.
We note that the Wigner function is 8 at $(r,p)=(0,0)$ regardless of the form of potential energy, because
\begin{eqnarray}
W(\vec{0},\vec{0})=\int d^3r^\prime \psi\bigg(\frac{\vec{r}^{~\prime}}{2}\bigg)\psi^*\bigg(-\frac{\vec{r}^{~\prime}}{2}\bigg)=8.
\label{t0}
\end{eqnarray}

One can see that the Wigner function at the large $m_D$ (or at high temperature) stretches in $r$-direction while it shrinks in $p$-direction, because the wavefunction extends in the coordinate space as temperature increases as shown in Fig.~\ref{waveft}.

\begin{figure} [h!]
    \includegraphics[width=8.5 cm]{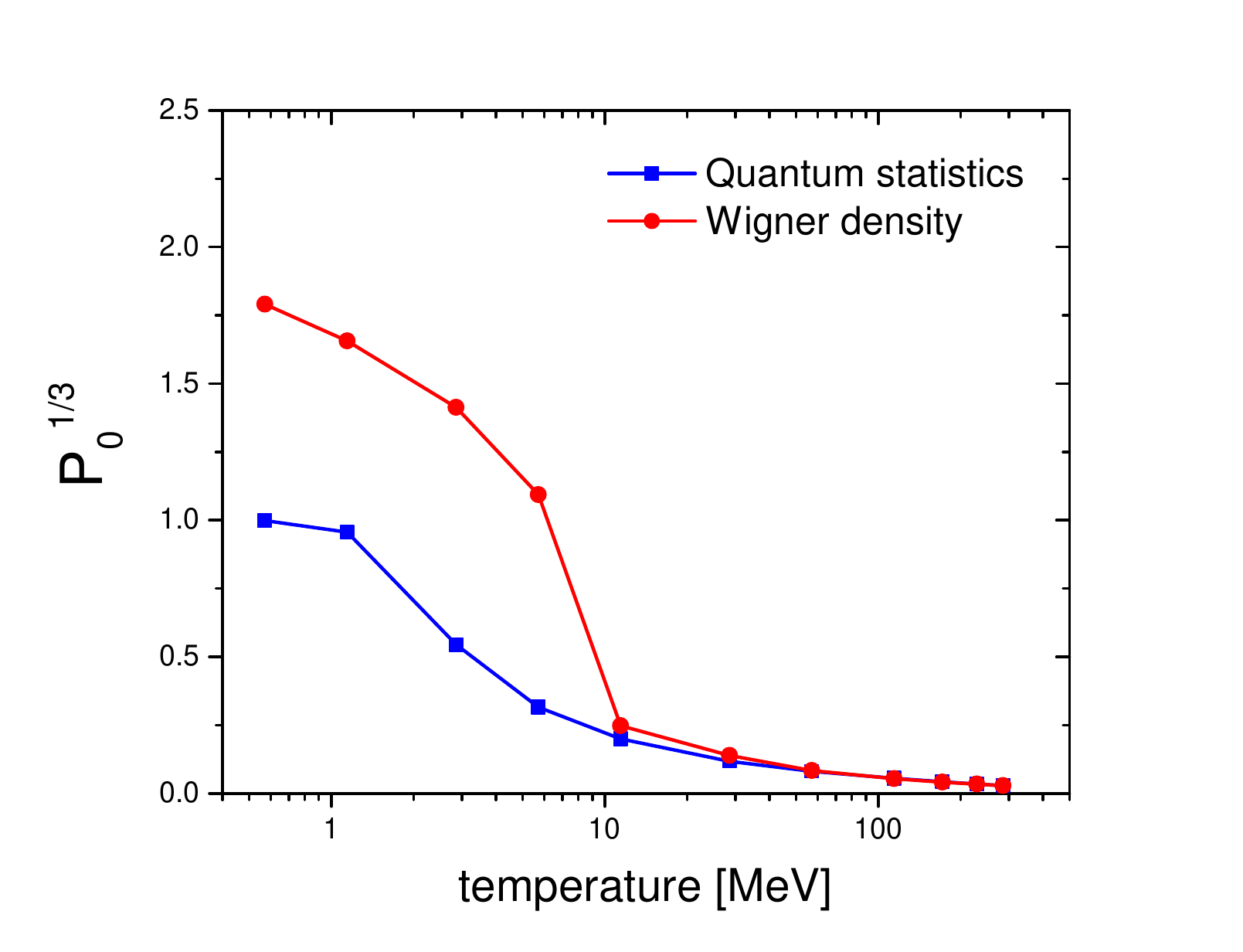}     
    \caption{The cube root probability of the ground state in the screened Coulomb potential as a function of temperature in quantum statistics and from the Wigner projection}
    \label{p0-coulomb}
\end{figure}

Fig.~\ref{p0-coulomb} shows the cube root probability for the ground state in quantum statistics and in the Wigner projection as a function of temperature.
The former is approximated by 
\begin{eqnarray}
P_0=\frac{e^{E/T}}{Z_{bound}+Z_{unbound}}\approx 
\frac{e^{E/T}}{e^{E/T}+V\{\mu T/(2\pi)\}^{3/2}}
\label{qs2}
\end{eqnarray}
where $E$ is the binding energy of the ground state in Fig.~\ref{waveft} and the excited bound states are neglected in $Z_{bound}$ for simplicity.
The ground state is however necessary for the convergence of $P_0=1$ in the limit of $T\rightarrow 0$.
The radial distance for the volume $V$ is taken to be five times average radius of wavefunction, which is insignificant at low temperature while important at high temperature where $Z_{unbound}$ is dominant over $Z_{bound}$. But for the purpose of comparison between quantum statistics and Wigner production, the volume is insignificant, because both approaches depend on the volume in Eqs.~(\ref{p0real}) and (\ref{qs2}).
We note that Fig.~\ref{p0-coulomb} is the cube root probability in comparison with Fig.~\ref{qm-wigner}, because the former is 3-dimensional while the latter 1-dimensional.
The cube root probability from the Wigner projection converges to 2 at $T=0$, where $\vec{r}, \vec{p}\rightarrow 0$, because of Eq.~(\ref{t0}). 
One can see that the behavior of the probability is similar in Figs.~\ref{qm-wigner} and \ref{p0-coulomb}.
It tells us that the Remler formalism is reliable unless temperature is too low.
But it begins to overestimate the formation of the bound state, as temperature decreases, due to the absence of quantum effect in the  particle distribution function with potential.

\section{quantum effects on spatial distribution}\label{quantum}

As mentioned in Sec.~\ref{1dsho}, particle position and momentum cannot simultaneously be specified in quantum mechanics.
Based on the uncertainty principle, once particle position is specified, particle momentum is unknown, and vice versa.
One can include the quantum effects simply by smearing the position and momentum of heavy (anti)quark~\cite{Han:2016uhh}.
In this study we focus on spatial smearing.
The smearing will be most effective near $r\sim 0$ where the potential rapidly changes.
So we simply assume that if the radial distance is smaller than $r_{min}$, the potential energy
is saturated: $V(r)=V(r_{min})$.
In other words, the potential is blurred in space with the resolution of $r_{min}$.
Now we return to Sec.~\ref{quarkonium} where quarkonium is formed by the free-energy potential.

\begin{figure} [h!]
    \includegraphics[width=8.5 cm]{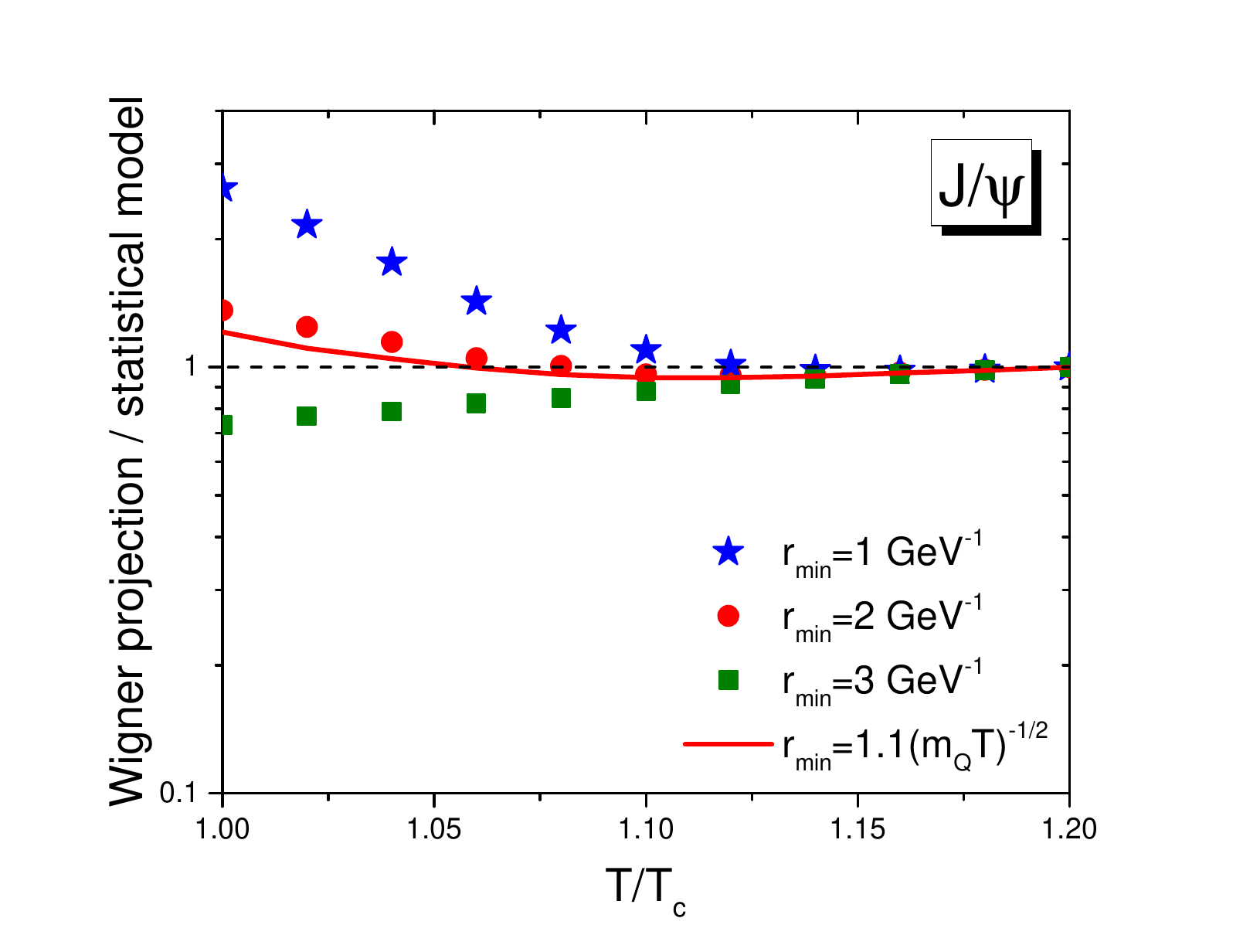}     
    \includegraphics[width=8.5 cm]{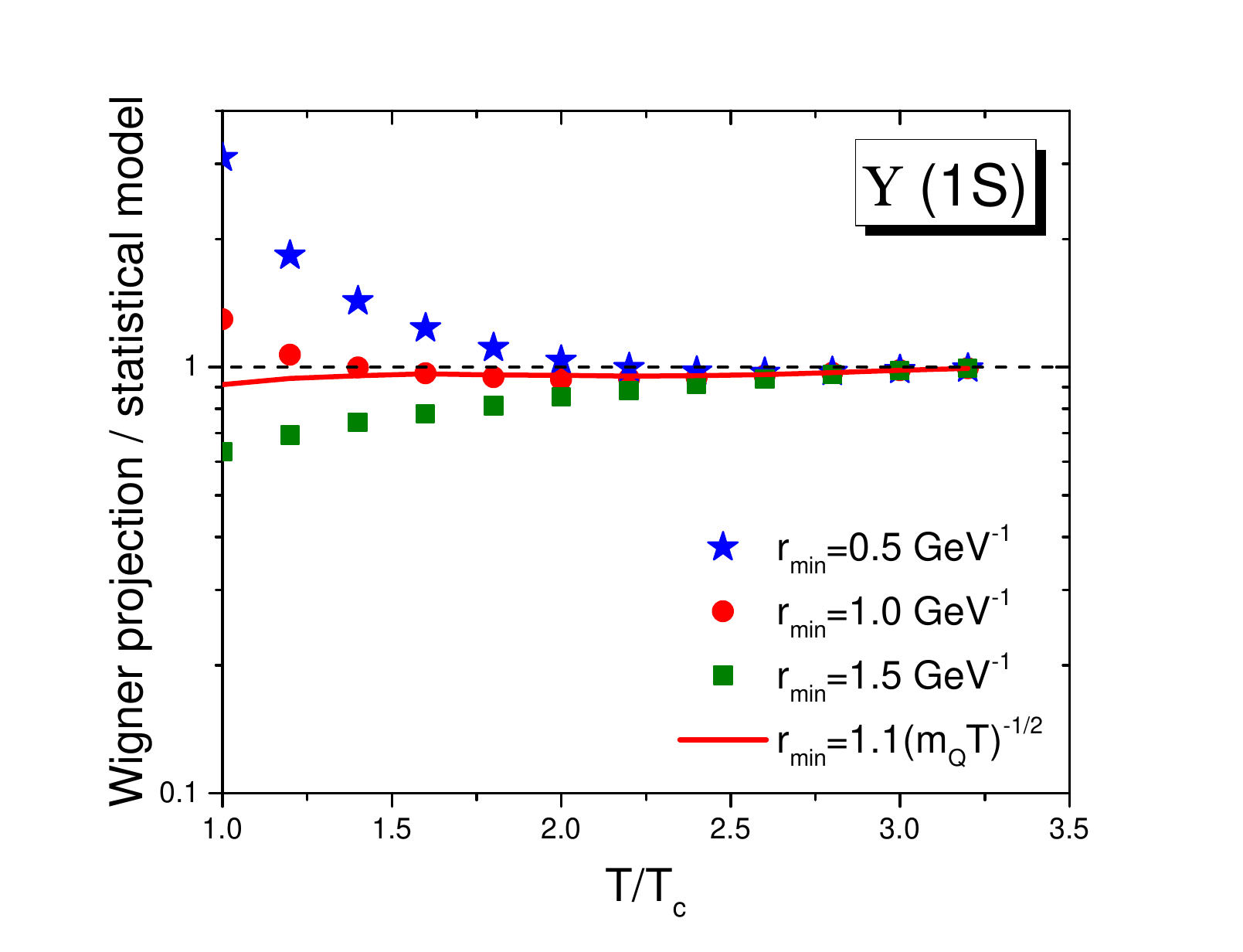}        
    \caption{The ratios of the number density of (upper) $J/\psi$ and of (lower) $\Upsilon (1S)$ from the Wigner projection to that from the statistical model for a couple of $r_{min}$ constant and proportional to the thermal wavelength ($\sim1/\sqrt{m_QT}$) as a function of rescaled temperature.}
    \label{ratio-new}
\end{figure}

The upper panel of Fig.~\ref{ratio-new} is the ratio of the number density of $J/\psi$ from the Wigner projection to that from the statistical model for $r_{min}=1,~2$ and 3 $\rm GeV^{-1}$ as a function of rescaled temperature.
Since the results from the Wigner function with the quarkonium wavefunction are quite similar to those from the gaussian Wigner function, as shown in Fig.~\ref{compare}, here we use the gaussian Wigner function. 
We note that $1/\mu$ in Eq.~(\ref{ratio}), the lower limit of $r$ integration, is not needed any more and the ratio is larger than in Fig.~\ref{compare} for $r_{min}=1~{\rm GeV^{-1}}$, because $1/\mu=2/m_Q\approx$ 1.3 $\rm GeV^{-1}$.
The ratio decreases with increasing $r_{min}$ and it crosses unit in the range of $r_{min}=2-3~ {\rm GeV^{-1}}$.
So the best $r_{min}$ to get the same results as in the statistical model will be there.
On the other hands, this kind of switching happens in the range of $r_{min}=1-1.5~ {\rm GeV^{-1}}$ for $\Upsilon (1S)$, as shown in the lower panel of Fig.~\ref{ratio-new}.
It implies that a heavier quark needs a smaller spatial smearing.
It is consistent with the thermal wavelength, $\lambda=\sqrt{2\pi/m_QT}$~\cite{Greiner:1995vqa}, though this form of wavelength is for free particle.
The red solid lines in Fig.~\ref{ratio-new} are the ratios for $r_{min}=1.1\sqrt{2\pi/m_QT}$.
It is interesting that the introduction of $r_{min}$ proportional to the thermal wavelength reproduces the statistical model results for both $J/\psi$ and $\Upsilon (1S)$.

\section{summary}\label{summary}

In our previous studies we have developed the Remler formalism for quarkonium production in thermal box, because the number of quarkonium must follow the thermal distribution in the box.
But if the binding energy of quarkonium is large, the Remler formalism or Wigner projection underestimates quarkonium production.

This underestimation can be cured by considering the potential between heavy quark pair, which attracts them to each other.
In this case the distance between heavy quark pair in color singlet decreases and Wigner projection is enhanced.

But we have found that the effect of heavy quark potential is too strong at low temperature, and the similar behavior is observed not only in the quarkonium production but also in the simple harmonic oscillator and the screened Coulomb potential.
The reason for the overestimation is the quantum effects which are more visible at low temperature.
By definition Wigner projection at $T=0$, where $r,p\rightarrow 0$, turns to $2^N$ with $N$ being the spatial dimension. 
However, position and momentum cannot be specified simultaneously in quantum mechanics due to the uncertainty principle.

In order to incorporate the quantum effect, the heavy quark potential is simply smeared by introducing the minimum radius which can physically be interpreted as the spatial resolution. 
We have found that $J/\psi$ needs larger smearing and $\Upsilon (1S)$ less smearing, and the smearing lengths are expressed in an unified form which is proportional to the thermal wavelength of heavy quark in QGP.

\begin{acknowledgments}
We appreciate Joerg Aichelin and Elena Bratkovskaya for their valuable discussions.
The work is supported by the Deutsche Forschungsgemeinschaft (DFG, German Research Foundation) through the grant CRC-TR 211 'Strong-interaction matter under extreme conditions' - Project number 315477589 - TRR 211 and the Helmholtz Research Academy Hessen for FAIR (HFHF). 
The computational resources have been provided by the LOEWE-Center for Scientific Computing and the "Green Cube" at GSI, Darmstadt and by the Center for Scientific Computing (CSC) of the Goethe University.
\end{acknowledgments}

\bibliography{main}

\end{document}